\documentclass[preprint,12pt,authoryear]{elsarticle}

\usepackage{amssymb}
\usepackage{amsmath}
\usepackage{graphicx}
\usepackage{booktabs}
\usepackage{array}
\usepackage{tabularx}
\usepackage{lineno}
\usepackage{subcaption}
\usepackage{float}
\usepackage{xurl}
\usepackage{hyperref}

\hypersetup{
    colorlinks=true,
    linkcolor=black,
    citecolor=blue,
    urlcolor=blue
}

\newcommand{\WSS}{\ensuremath{\mathrm{WSS}}}

\begin{document}

\begin{frontmatter}

\title{Multi-Granularity Conformal Prediction for Reliable Neural-Operator Automotive Aerodynamic Surrogates}

\author[label1]{Chundong Jia}
\author[label2]{Chao Xia\corref{cor1}}
\ead{chao.xia@chalmers.se}
\cortext[cor1]{Corresponding author.}
\author[label2]{Alexey Vdovin}
\author[label3]{Qing Jia}
\author[label2]{Simone Sebben}
\author[label3]{Zhigang Yang}

\affiliation[label1]{organization={Department of Mathematics},
addressline={Karlsruhe Institute of Technology}, city={Karlsruhe}, postcode={76131}, country={Germany}}
\affiliation[label2]{organization={Department of Mechanical Engineering},
addressline={Chalmers University of Technology}, city={Gothenburg}, postcode={412 96}, country={Sweden}}
\affiliation[label3]{organization={College of Automotive and Energy Engineering},
addressline={Tongji University}, city={Shanghai}, postcode={201804}, country={China}}

\begin{abstract}
High-fidelity computational fluid dynamics (CFD) provides detailed aerodynamic quantities for vehicle design, but its cost limits rapid design iteration. Neural-operator surrogates reduce this cost, yet their deterministic predictions provide little information about when a predicted vehicle geometry or surface region should be trusted. This study develops a conformal-prediction framework for reliability-aware automotive aerodynamic surrogate modeling on the DrivAerML dataset. GeoTransolver is used as the main backbone, and Transolver is used to examine transfer across neural-operator architectures. For scalar drag coefficient prediction, conformalized quantile regression constructs calibrated case-level intervals. For surface pressure and wall shear stress (WSS), point prediction is combined with residual-scale estimation and residual-normalized conformal calibration to obtain spatially adaptive intervals. Global absolute, point-adaptive normalized, and case-wise normalized calibration are compared under ordinary split and cross-validation-assisted out-of-fold protocols.
All experiments target 90\% nominal coverage. Conformal calibration corrects the under-coverage of raw drag-coefficient quantile intervals, and out-of-fold score aggregation reduces the Monte Carlo coverage standard deviation from 10.41 to 3.10 percentage points. For surface fields, point-adaptive normalized calibration gives the narrowest near-nominal intervals, reducing mean width by 22.68\% for pressure and by 25.35\%--27.09\% for WSS under the out-of-fold protocol. Case-wise normalized calibration is more conservative but improves vehicle-level reliability. Smoothness regularization reduces the residual-scale local-variation score by 74.29\% and lowers interval widths without material coverage loss. The resulting framework turns deterministic neural-operator predictions into calibrated reliability indicators for prioritizing uncertain vehicle geometries and surface regions in follow-up CFD verification.
\end{abstract}



\end{frontmatter}


\section{Introduction}
\label{sec:introduction}

Aerodynamic performance is a key factor in vehicle design because it affects energy consumption, driving stability, and thermal management. High-fidelity computational fluid dynamics (CFD) can provide detailed drag coefficient, surface pressure, and wall shear stress (WSS) information for complex vehicle geometries, but its computational cost limits rapid design exploration and optimization.

Data-driven aerodynamic surrogate models offer a practical way to reduce this cost. Public datasets such as DrivAerNet and DrivAerNet++ have supported large-scale data-driven aerodynamic prediction for vehicle geometries~\citep{elrefaie2024drivaernet,elrefaie2024drivaernetpp}, while DrivAerML further provides high-fidelity CFD data for realistic automotive aerodynamic surrogate modeling~\citep{ashton2024drivaerml}. Recent geometry-aware neural operators and transformer-based surrogates have improved aerodynamic prediction on complex vehicle surfaces~\citep{ranade2025domino,wu2024transolver,alkin2025abupt,adams2025geotransolver}, making them attractive for fast aerodynamic screening.

However, most aerodynamic surrogates are still used as deterministic predictors. They provide a single scalar value or surface field but do not indicate whether the prediction is reliable. This is problematic in engineering applications because a model with low average error may still fail on specific vehicle geometries or local surface regions, such as wheels, sharp edges, underbody structures, or wake-influenced areas. Without calibrated uncertainty information, it is difficult to decide which predictions can be trusted and which cases require additional CFD verification.

Conformal prediction provides a model-agnostic way to construct calibrated prediction intervals with a finite-sample coverage interpretation under exchangeability assumptions~\citep{vovk2005algorithmic,shafer2008tutorial,angelopoulos2021gentle}. Recent studies have extended conformal calibration to operator learning, DeepONet regression, high-dimensional scientific surrogates, and dynamical-system rollouts~\citep{ma2024calibrated,moya2025conformalized,gopakumar2026uqsurrogate,liang2024dynamiccp}. Nevertheless, applying conformal prediction to automotive aerodynamic neural operators remains nontrivial. Scalar drag coefficient prediction and surface-field prediction have different output dimensions, error structures, and calibration requirements. In particular, pressure and wall shear stress errors can vary across vehicle samples, physical channels, and local surface regions.  Therefore, the key problem is not simply to attach conformal prediction to a trained surrogate, but to design output-specific calibration strategies that reflect the heterogeneous aerodynamic outputs and engineering reliability requirements.

This work develops a task-adaptive conformal reliability framework for neural-operator-based automotive aerodynamic surrogates on the DrivAerML dataset. GeoTransolver is used as the main surrogate backbone, and Transolver is used to examine transferability across neural-operator architectures. For scalar drag coefficient prediction, conformalized quantile regression is used to construct calibrated case-level intervals. For surface pressure and wall shear stress, point prediction is combined with learned residual-scale estimation and residual-normalized conformal calibration to produce spatially adaptive intervals.

The main contributions are as follows:
\begin{itemize}
    \item We develop a task-adaptive conformal reliability framework for neural-operator automotive aerodynamic surrogates, unifying vehicle-level scalar drag prediction and surface-field prediction within one calibrated reliability workflow.
    \item We adapt residual-normalized conformal prediction to pressure and wall shear stress prediction by learning pointwise residual-scale fields, enabling spatially adaptive uncertainty intervals.
    \item We introduce a local-neighborhood smoothness regularization for the residual-scale field to improve spatial coherence and reduce nonphysical local jumps.
    \item We compare global absolute, point-adaptive normalized, and case-wise normalized calibration under ordinary split and CV-assisted out-of-fold protocols, revealing trade-offs between interval width, coverage, and vehicle-level reliability.
    \item We evaluate the same calibration pipeline with GeoTransolver and Transolver, showing that the framework can transfer across neural-operator backbones and support reliability-aware CFD verification.
\end{itemize}
\section{Related Work}
\label{sec:related_work}

\subsection{Automotive aerodynamic datasets and surrogate prediction}
\label{subsec:cfd_aero_prediction}

CFD has long been used in automotive aerodynamic design because it provides both global coefficients and local flow-field quantities. Drag coefficient is directly related to aerodynamic efficiency, while surface pressure and wall shear stress describe local loading, near-wall flow behavior, and separation-related effects. However, the high cost of CFD has motivated the development of data-driven surrogate models for rapid aerodynamic prediction.

Public datasets have accelerated this development. The DrivAer model provides a standardized realistic vehicle geometry for reproducible aerodynamic studies~\citep{heft2012drivaer}. DrivAerNet and DrivAerNet++ provide large-scale vehicle geometries with CFD-derived aerodynamic labels for learning-based prediction and design studies~\citep{elrefaie2024drivaernet,elrefaie2024drivaernetpp}. DrivAerML further provides high-fidelity CFD data for parametrically morphed DrivAer vehicles, including drag coefficient, surface pressure, and wall shear stress~\citep{ashton2024drivaerml}. These datasets make it possible to evaluate surrogate models on realistic automotive geometries and compare their behavior across scalar and field-level prediction tasks.

Earlier aerodynamic surrogates mainly focused on coefficient-level prediction, whereas recent work increasingly targets field-level prediction on meshes, point clouds, or arbitrary query locations. Field-level prediction is more informative for engineering analysis because it preserves spatial information about pressure distribution, wall shear stress, and local flow features. Nevertheless, most studies still report deterministic error metrics such as MAE, RMSE, relative L1/L2 error, or \(R^2\). These metrics are necessary for accuracy evaluation, but they do not quantify the reliability of individual vehicle cases or local surface regions.

\subsection{Neural operators for aerodynamic surrogate modeling}
\label{subsec:neural_operator_surrogate}

Neural operators, represented by DeepONet and Fourier neural operator, learn mappings from input functions, parameters, or geometries to output fields~\citep{lu2021deeponet,li2021fno}. They have been applied to aerodynamic flow-field prediction, shape optimization, and other full-field engineering tasks, while geometry-adaptive variants extend operator learning to varying domains such as cylinder and airfoil flows~\citep{shukla2024deep,he2024sequential,lee2026geometry}. This makes them well suited to learning geometry-dependent mappings from vehicle surfaces to aerodynamic quantities.

Recent neural-operator and transformer-based methods have focused on complex geometries and irregular domains. GINO combines geometric processing with Fourier neural operators for large-scale three-dimensional PDEs~\citep{li2023gino}. DoMINO uses decomposable multi-scale operators for engineering simulations and has been applied to DrivAerML tasks~\citep{ranade2025domino}. Transolver introduces physics attention over learned physical states for PDEs on general geometries~\citep{wu2024transolver}. GAOT combines geometry embeddings, graph-based encoders and decoders, and transformer processors for PDEs on arbitrary domains~\citep{wen2025gaot}. AB-UPT targets high-fidelity automotive CFD surrogates at industrial mesh scales~\citep{alkin2025abupt}. SATO and SMART further explore transformer-based aerodynamic prediction on arbitrary vehicle geometries and raw point-cloud inputs~\citep{yang2026sato,hagnberger2026smart}. GeoTransolver extends the Transolver family with multi-scale geometry-aware context and has been benchmarked on DrivAerML and related datasets~\citep{adams2025geotransolver}.

These methods provide strong deterministic backbones for aerodynamic surrogate modeling. However, their predictions are usually evaluated mainly by point-prediction accuracy. The reliability of their outputs at the vehicle level and surface-region level remains insufficiently characterized, especially for pressure and multi-component wall shear stress prediction.

\subsection{Uncertainty quantification and conformal prediction}
\label{subsec:uq_conformal_prediction}

Uncertainty quantification has been widely studied for deep learning models. Bayesian neural networks require approximate posterior inference and sampling~\citep{blundell2015weight}, Monte Carlo dropout relies on repeated stochastic forward passes~\citep{gal2016dropout}, and deep ensembles require multiple independently trained models~\citep{lakshminarayanan2017deepensembles}. These costs are particularly relevant for automotive aerodynamic surrogates with many surface points and physical output channels.

Uncertainty estimation has also been incorporated into operator-learning architectures. For example, VB-DeepONet applies variational Bayesian inference to operator regression~\citep{garg2023vbdeeponet}. However, such methods modify the training framework and do not directly provide the finite-sample coverage interpretation of conformal prediction.

Conformal prediction provides a post-hoc, model-agnostic approach to constructing prediction sets by calibrating nonconformity scores on held-out data, with finite-sample coverage under exchangeability assumptions~\citep{vovk2005algorithmic,shafer2008tutorial,angelopoulos2021gentle}. Conformalized quantile regression further calibrates learned quantile intervals~\citep{romano2019cqr}. Prediction intervals and conformal methods have been applied to transportation, renewable-energy forecasting, and industrial deep-learning systems~\citep{mazloumi2011prediction,nourani2025uncertainty,dossantos2025robust,mehdiyev2025conformal}. Because calibration is performed after model training, conformal prediction is well suited to neural-operator surrogates that are costly to retrain or replicate.

Recent studies have extended conformal prediction to scientific operator learning and high-dimensional surrogate modeling. Function-space conformal prediction has been evaluated on PDEs and vehicle surface-pressure prediction~\citep{ma2024calibrated}, while conformalized DeepONet provides distribution-free intervals for operator regression~\citep{moya2025conformalized}. Other work has considered high-dimensional surrogate outputs, dynamical-system rollouts, and aerodynamic or neural-fluid applications~\citep{gopakumar2026uqsurrogate,liang2024dynamiccp,nietocentenero2025multifidelity}. Cross-validation-based methods such as jackknife+ further motivate multi-fold calibration for improved data use and reduced split dependence~\citep{barber2021jackknife}.

Despite these developments, conformal calibration for neural-operator automotive aerodynamic surrogates remains underexplored. It is unclear how scalar drag coefficient, surface pressure, and three-component wall shear stress should be calibrated within a unified framework, and how calibration granularity, OOF score aggregation, residual-scale smoothness, and backbone replacement affect interval efficiency and case-level reliability. This work addresses these questions on the DrivAerML benchmark.

\section{Methodology}
\label{sec:methodology}

This section defines a task-adaptive conformal reliability framework for neural-operator-based automotive aerodynamic surrogates. The framework is built around three design choices: output-specific interval construction, residual-scale-based spatial adaptivity, and calibration granularity for vehicle-level reliability. For scalar drag coefficient prediction, an interval-oriented predictor is followed by conformal calibration. For surface pressure and wall shear stress, point prediction is combined with residual-scale estimation and residual-normalized conformal calibration. This separation is necessary because \(C_d\) is a case-level scalar quantity, whereas pressure and wall shear stress are surface-distributed fields whose errors may vary across points, physical channels, and vehicle samples.

\subsection{Problem formulation}
\label{subsec:problem_formulation}

A vehicle surface is represented by a set of sampled surface points:
\begin{equation}
\mathcal{X}
=
\left\{
(\mathbf{x}_i,\mathbf{n}_i,\mathbf{a}_i)
\right\}_{i=1}^{N}.
\end{equation}
Here, \(\mathbf{x}_i \in \mathbb{R}^{3}\) denotes the coordinate of the \(i\)-th surface point, \(\mathbf{n}_i \in \mathbb{R}^{3}\) denotes the corresponding surface normal, and \(\mathbf{a}_i\) denotes additional geometric or auxiliary features if available. The number of sampled surface points is denoted by \(N\).

This work considers two aerodynamic prediction settings. The first is scalar drag coefficient prediction:
\begin{equation}
C_d \in \mathbb{R}.
\end{equation}
The second is surface-field prediction. For surface pressure, the target field is
\begin{equation}
\mathbf{y}^{p}
=
\left\{
p_i
\right\}_{i=1}^{N},
\end{equation}
and for three-dimensional wall shear stress, the target field is
\begin{equation}
\mathbf{y}^{\tau}
=
\left\{
(\tau_{x,i},\tau_{y,i},\tau_{z,i})
\right\}_{i=1}^{N}.
\end{equation}
Here, \(p_i\) denotes the pressure value at the \(i\)-th surface point, and \((\tau_{x,i},\tau_{y,i},\tau_{z,i})\) denotes the three wall shear stress components at the same point.

For each prediction setting, the objective is to obtain both a point prediction \(\hat{y}\) and a prediction interval:
\begin{equation}
\mathcal{I}(\mathcal{X})
=
[
\hat{y}^{L}(\mathcal{X}),
\hat{y}^{U}(\mathcal{X})
].
\end{equation}
For surface fields, this notation is applied point-wise and component-wise. Unless otherwise stated, all intervals target \(90\%\) nominal coverage, corresponding to a miscoverage rate \(\alpha=0.10\). For scalar drag coefficient prediction, one calibrated interval is constructed for each vehicle geometry. For surface-field prediction, intervals are constructed for pressure or wall shear stress values at surface points.

For a calibration score set \(\mathcal{S}=\{s_j\}_{j=1}^{m}\), the conformal quantile is computed as the \(k\)-th order statistic:
\begin{equation}
Q_{1-\alpha}(\mathcal{S})
=
s_{(k)},
\qquad
k=
\left\lceil
(m+1)(1-\alpha)
\right\rceil ,
\end{equation}
where \(s_{(k)}\) denotes the \(k\)-th smallest value among the sorted calibration scores. In implementation, \(k\) is clipped at \(m\) if necessary. Under exchangeability, this finite-sample correction gives the usual marginal conformal coverage guarantee, with the finite-sample discreteness term described in standard conformal prediction theory~\citep{vovk2005algorithmic,shafer2008tutorial,angelopoulos2021gentle}. When asymmetric scalar calibration is used, \(Q_{1-\alpha/2}(\cdot)\) is computed analogously.

\subsection{Neural operator surrogate backbones}
\label{subsec:surrogate_backbones}

GeoTransolver is used as the main surrogate backbone, and Transolver is used as an additional backbone for transferability evaluation. Both models are treated as existing neural-operator aerodynamic surrogate models rather than newly proposed architectures. For each task, the backbone maps the input surface representation \(\mathcal{X}\) to geometry-dependent latent features, while the prediction head and conformal calibration procedure are defined according to the output type. The scalar task uses interval-related outputs, whereas the surface-field task predicts both field values and local residual scales.

\subsection{Conformal prediction for scalar drag coefficient estimation}
\label{subsec:cd_conformal}

The drag coefficient task is formulated as a case-level scalar prediction problem. Because \(C_d\) is defined for the entire vehicle geometry rather than for individual surface points, the point-wise representations produced by the neural-operator backbone must be converted into a vehicle-level representation before scalar prediction. GeoTransolver first maps the sampled surface points to point-wise latent features:
\begin{equation}
\mathbf{H}
=
f_{\theta}^{geo}(\mathcal{X})
=
\left\{
\mathbf{h}_i
\right\}_{i=1}^{N},
\qquad
\mathbf{h}_i \in \mathbb{R}^{d_h}.
\end{equation}
Here, \(\mathbf{h}_i\) denotes the latent feature of the \(i\)-th point, and \(d_h\) denotes the hidden feature dimension.

To adapt GeoTransolver to scalar drag coefficient regression, a structured readout module aggregates the point-wise features into a vehicle-level feature. The readout contains global statistical pooling, region-wise statistical pooling, and global geometry descriptors.

First, global statistical pooling is applied over all sampled surface points:
\begin{equation}
\mathbf{z}_{global}
=
\operatorname{Concat}
\left(
\operatorname{Mean}_{i=1}^{N}(\mathbf{h}_i),
\operatorname{Max}_{i=1}^{N}(\mathbf{h}_i),
\operatorname{Std}_{i=1}^{N}(\mathbf{h}_i)
\right).
\end{equation}

Second, region-wise pooling is performed over predefined vehicle regions. The surface points are divided into front, middle, rear, and underbody regions according to their normalized spatial coordinates. Let \(\mathcal{R}\) denote this region set. For each region \(r\in\mathcal{R}\), masked statistical pooling is applied:
\begin{equation}
\mathbf{z}_{r}
=
\operatorname{Concat}
\left(
\operatorname{Mean}_{i\in r}(\mathbf{h}_i),
\operatorname{Max}_{i\in r}(\mathbf{h}_i),
\operatorname{Std}_{i\in r}(\mathbf{h}_i)
\right).
\end{equation}

Third, global geometry descriptors provide case-level geometric information and are denoted by \(\mathbf{z}_{geo}\). The final vehicle-level representation for drag coefficient prediction is
\begin{equation}
\mathbf{z}_{C_d}
=
\operatorname{Concat}
\left(
\mathbf{z}_{global},
\left\{
\mathbf{z}_{r}
\right\}_{r\in\mathcal{R}},
\mathbf{z}_{geo}
\right).
\end{equation}

\begin{figure}[t]
    \centering
    \includegraphics[width=\textwidth]{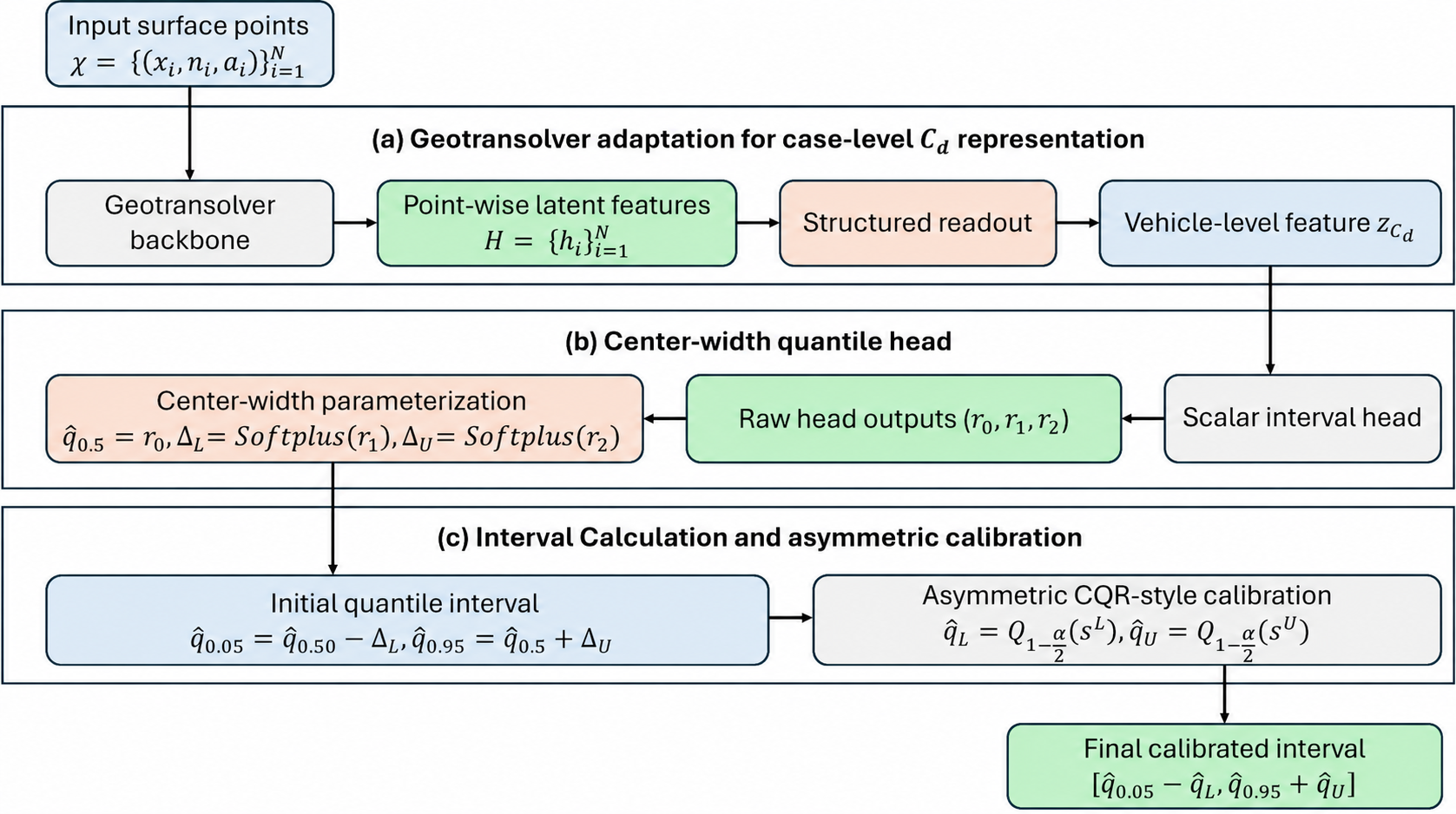}
    \caption{Scalar drag coefficient interval prediction with connected readout and calibration stages.
    The GeoTransolver backbone first produces point-wise latent features, which are aggregated into a vehicle-level feature \(\mathbf{z}_{C_d}\).
    This feature is then passed to the center-width quantile head and asymmetric conformal calibration module.}
    \label{fig:cd_interval_method}
\end{figure}

This structured readout converts the point-wise neural-operator representation into a scalar-oriented representation for drag coefficient prediction. The scalar readout, center-width quantile head, and asymmetric conformal calibration workflow are summarized in Fig.~\ref{fig:cd_interval_method}.

The scalar prediction head predicts the \(0.05\), \(0.50\), and \(0.95\) quantile levels:
\begin{equation}
(r_0,r_1,r_2)
=
h_{\theta}^{C_d}
\left(
\mathbf{z}_{C_d}
\right).
\end{equation}
Rather than treating these outputs as independent quantiles, the model uses a center-width parameterization:
\begin{equation}
\hat{q}_{0.50}(\mathcal{X})
=
r_0,
\end{equation}
\begin{equation}
\Delta_L(\mathcal{X})
=
\operatorname{Softplus}(r_1),
\qquad
\Delta_U(\mathcal{X})
=
\operatorname{Softplus}(r_2),
\end{equation}
\begin{equation}
\hat{q}_{0.05}(\mathcal{X})
=
\hat{q}_{0.50}(\mathcal{X})
-
\Delta_L(\mathcal{X}),
\qquad
\hat{q}_{0.95}(\mathcal{X})
=
\hat{q}_{0.50}(\mathcal{X})
+
\Delta_U(\mathcal{X}).
\end{equation}
This center-width parameterization enforces
\begin{equation}
\hat{q}_{0.05}(\mathcal{X})
\leq
\hat{q}_{0.50}(\mathcal{X})
\leq
\hat{q}_{0.95}(\mathcal{X}),
\end{equation}
thereby avoiding quantile crossing while allowing asymmetric lower and upper widths.

The quantile outputs are trained using the pinball loss. For a scalar target \(y\), a predicted \(\rho\)-quantile \(\hat{q}_{\rho}\), and \(\rho\in\{0.05,0.50,0.95\}\), this loss is
\begin{equation}
\mathcal{L}_{\rho}
\left(
y,
\hat{q}_{\rho}
\right)
=
\max
\left\{
\rho
\left(
y-\hat{q}_{\rho}
\right),
(\rho-1)
\left(
y-\hat{q}_{\rho}
\right)
\right\}.
\end{equation}
The training loss for drag coefficient prediction is
\begin{equation}
\mathcal{L}_{C_d}
=
\frac{1}{3}
\sum_{\rho\in\{0.05,0.50,0.95\}}
\mathcal{L}_{\rho}
\left(
C_d,
\hat{q}_{\rho}
\right).
\end{equation}
The median output \(\hat{q}_{0.50}\) is used as the point prediction, while \(\hat{q}_{0.05}\) and \(\hat{q}_{0.95}\) define the initial nominal interval.

After training, conformal calibration is applied to a held-out calibration set:
\begin{equation}
\mathcal{D}_{cal}
=
\left\{
\left(
\mathcal{X}_j,
C_{d,j}
\right)
\right\}_{j=1}^{m}.
\end{equation}
Because the lower and upper tails may have different calibration errors, this work uses an asymmetric CQR-style calibration. The lower and upper nonconformity scores are defined using positive parts:
\begin{equation}
s_j^{L}
=
\max
\left\{
\hat{q}_{0.05}
\left(
\mathcal{X}_j
\right)
-
C_{d,j},
0
\right\},
\qquad
s_j^{U}
=
\max
\left\{
C_{d,j}
-
\hat{q}_{0.95}
\left(
\mathcal{X}_j
\right),
0
\right\}.
\end{equation}
These positive-part scores measure the required downward and upward expansions of the initial interval.

The positive-part formulation only preserves or expands the initial interval. This conservative design is used because the calibrated interval serves as a reliability indicator for deciding whether additional CFD verification is needed; avoiding under-coverage is therefore prioritized over post-hoc interval shrinking.

For a target miscoverage level \(\alpha\), the lower and upper calibration offsets are computed separately:
\begin{equation}
\hat{q}_{L}
=
Q_{1-\alpha/2}
\left(
\left\{
s_j^{L}
\right\}_{j=1}^{m}
\right),
\qquad
\hat{q}_{U}
=
Q_{1-\alpha/2}
\left(
\left\{
s_j^{U}
\right\}_{j=1}^{m}
\right).
\end{equation}
The calibrated prediction interval for a test vehicle \(\mathcal{X}_{test}\) is then
\begin{equation}
\mathcal{I}_{C_d}
\left(
\mathcal{X}_{test}
\right)
=
\left[
\hat{q}_{0.05}
\left(
\mathcal{X}_{test}
\right)
-
\hat{q}_{L},
\quad
\hat{q}_{0.95}
\left(
\mathcal{X}_{test}
\right)
+
\hat{q}_{U}
\right].
\end{equation}
The two calibration offsets are not forced to be identical, so the calibrated interval can account for asymmetric lower- and upper-tail errors.

\subsection{Residual-normalized conformal prediction for surface fields}
\label{subsec:field_conformal}

Surface pressure and wall shear stress are treated as surface-field prediction tasks. Unlike the scalar drag coefficient task, the surface-field model predicts a point estimate and a residual scale field rather than quantile-regression outputs. For a surface-field target \(\mathbf{y}\), the model outputs
\begin{equation}
\hat{\mathbf{y}}
=
f_{\theta}^{mean}(\mathcal{X}),
\qquad
\hat{\boldsymbol{\sigma}}
=
f_{\theta}^{scale}(\mathcal{X}),
\end{equation}
where \(\hat{\mathbf{y}}\) denotes the predicted pressure or wall shear stress field, and \(\hat{\boldsymbol{\sigma}}\) denotes the estimated residual scale. In implementation, the residual-scale branch takes local geometric information and the detached mean prediction as input. The predicted scale is constrained to be positive through a Softplus transformation with a small lower bound.

For a surface point \(i\) and component \(c\), the absolute residual used as the scale target is
\begin{equation}
r_{i,c}
=
\left|
y_{i,c}
-
\hat{y}_{i,c}
\right|.
\end{equation}
To avoid unstable scale targets near zero, the residual target is lower-bounded by a small floor value:
\begin{equation}
\tilde{r}_{i,c}
=
\max
\left(
r_{i,c},
r_{\min}
\right).
\end{equation}
The residual-scale branch is trained in log space using a smooth L1 loss:
\begin{equation}
\mathcal{L}_{scale}
=
\frac{1}{NC}
\sum_{i=1}^{N}
\sum_{c=1}^{C}
\operatorname{SmoothL1}
\left(
\log
\left(
\hat{\sigma}_{i,c}
+
\epsilon
\right),
\log
\left(
\mathrm{sg}
\left(
\tilde{r}_{i,c}
\right)
+
\epsilon
\right)
\right),
\end{equation}
where \(C=4\) corresponds to \((p,\tau_x,\tau_y,\tau_z)\), \(\epsilon>0\) is used for numerical stability, and \(\mathrm{sg}(\cdot)\) stops gradients through the residual target.

The full training objective for the surface-field model is
\begin{equation}
\mathcal{L}_{field}
=
\mathcal{L}_{mean}
+
\lambda_{\mathrm{scale}}
\mathcal{L}_{scale}
+
\lambda_{\mathrm{smooth}}
\mathcal{L}_{smooth},
\end{equation}
where \(\mathcal{L}_{mean}\) is the prediction loss for the surface-field value, \(\mathcal{L}_{scale}\) trains the residual-scale branch, and \(\mathcal{L}_{smooth}\) is the spatial smoothness regularization term introduced in Section~\ref{subsec:smoothness_regularization}. The values of \(\lambda_{\mathrm{scale}}\) and \(\lambda_{\mathrm{smooth}}\) are specified in Table~\ref{tab:model_configurations}.

Given a calibration set, the residual-normalized conformal score is computed as
\begin{equation}
s_{b,i,c}^{field}
=
\frac{
\left|
y_{b,i,c}
-
\hat{y}_{b,i,c}
\right|
}{
\hat{\sigma}_{b,i,c}
+
\epsilon
},
\end{equation}
where \(b\) indexes the vehicle sample, \(i\) indexes the surface point, and \(c\) indexes the physical component. This score measures the prediction error after normalization by the learned local residual scale. It is used by the normalized calibration strategies described in Section~\ref{subsec:calibration_granularity}. A global absolute calibration strategy based directly on absolute residuals is also retained as a non-adaptive baseline.

\subsection{Calibration granularity for surface-field uncertainty}
\label{subsec:calibration_granularity}

For surface-field prediction, this work compares three calibration strategies: global absolute calibration, point-adaptive normalized calibration, and case-wise normalized calibration. The residual-normalized score \(s_{b,i,c}^{field}\) is defined in Section~\ref{subsec:field_conformal}. Because the surface-field model predicts four output channels, namely pressure and the three wall shear stress components \((p,\tau_x,\tau_y,\tau_z)\), conformal quantiles are computed separately for each channel.

The three strategies represent increasing levels of adaptivity. Global absolute calibration provides a channel-wise non-adaptive baseline. Point-adaptive normalized calibration produces spatially varying intervals through the learned residual scale. Because surface points on the same vehicle are spatially correlated, point-pooled coverage alone may not fully reflect reliability at the vehicle-case level. Case-wise normalized calibration therefore aggregates normalized scores within each vehicle sample to improve empirical vehicle-level reliability.

\begin{figure}[htbp]
    \centering
    \includegraphics[width=\textwidth]{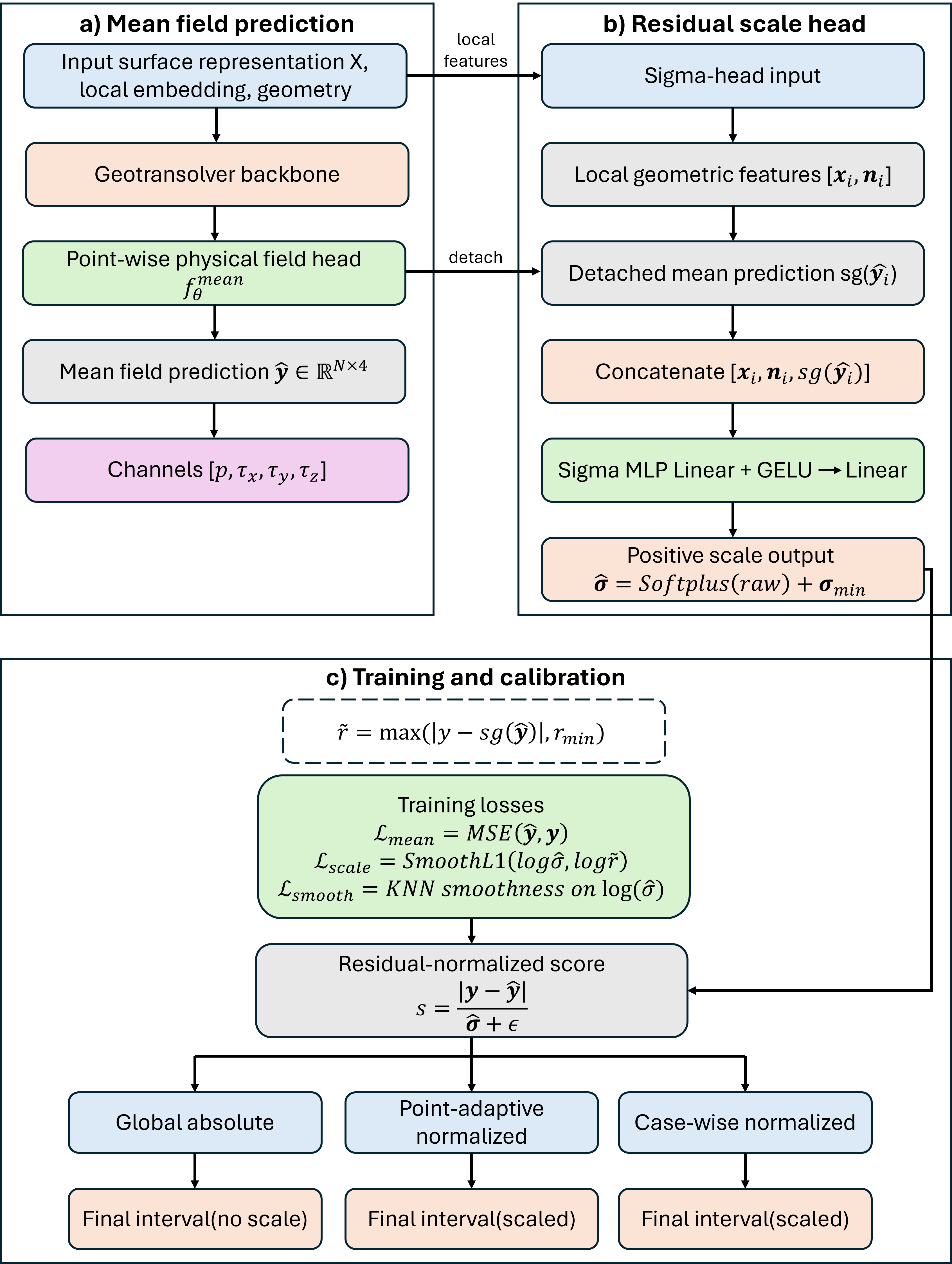}
    \caption{Surface-field prediction and calibration workflow. The model estimates both the mean field \(\hat{\mathbf{y}}\) and residual scale \(\hat{\boldsymbol{\sigma}}\), followed by global absolute, point-adaptive normalized, and case-wise normalized calibration.}
    \label{fig:surface_field_calibration}
\end{figure}

The overall surface-field prediction and calibration procedure is illustrated in Fig.~\ref{fig:surface_field_calibration}.

In global absolute calibration, the calibration score is the absolute residual. For calibration sample \(b\), surface point \(i\), and component \(c\), let
\begin{equation}
a_{b,i,c}
=
\left|
y_{b,i,c}
-
\hat{y}_{b,i,c}
\right|.
\end{equation}
All absolute residuals from the same output channel are pooled over calibration samples and surface points:
\begin{equation}
\hat{q}_{c}^{global}
=
Q_{1-\alpha}
\left(
\left\{
a_{b,i,c}
\right\}_{b,i}
\right).
\end{equation}
The resulting prediction interval is
\begin{equation}
\mathcal{I}_{i,c}^{global}
=
\left[
\hat{y}_{i,c}
-
\hat{q}_{c}^{global},
\quad
\hat{y}_{i,c}
+
\hat{q}_{c}^{global}
\right].
\end{equation}

In point-adaptive normalized calibration, the residual-normalized scores from the same output channel are pooled over calibration samples and surface points:
\begin{equation}
\hat{q}_{c}^{point}
=
Q_{1-\alpha}
\left(
\left\{
s_{b,i,c}^{field}
\right\}_{b,i}
\right).
\end{equation}
The calibrated interval is constructed as
\begin{equation}
\mathcal{I}_{i,c}^{point}
=
\left[
\hat{y}_{i,c}
-
\hat{q}_{c}^{point}
\hat{\sigma}_{i,c},
\quad
\hat{y}_{i,c}
+
\hat{q}_{c}^{point}
\hat{\sigma}_{i,c}
\right].
\end{equation}

In case-wise normalized calibration, the point-level normalized scores are first reduced within each vehicle sample. For each output channel \(c\), the case-level score is defined as
\begin{equation}
s_{b,c}^{case}
=
A_{case}
\left(
\left\{
s_{b,i,c}^{field}
\right\}_{i=1}^{N}
\right),
\end{equation}
where \(A_{case}(\cdot)\) denotes a within-case aggregation operator. In this work, \(A_{case}(\cdot)\) is implemented as the within-case empirical quantile of the point-level normalized scores:
\begin{equation}
s_{b,c}^{case}
=
Q_{1-\alpha}
\left(
\left\{
s_{b,i,c}^{field}
\right\}_{i=1}^{N}
\right).
\end{equation}
This choice summarizes the high-error portion of each vehicle sample while avoiding the excessive conservativeness of a maximum-based aggregation. The case-wise conformal quantile is then computed over calibration samples:
\begin{equation}
\hat{q}_{c}^{case}
=
Q_{1-\alpha}
\left(
\left\{
s_{b,c}^{case}
\right\}_{b}
\right).
\end{equation}
The resulting prediction interval is
\begin{equation}
\mathcal{I}_{i,c}^{case}
=
\left[
\hat{y}_{i,c}
-
\hat{q}_{c}^{case}
\hat{\sigma}_{i,c},
\quad
\hat{y}_{i,c}
+
\hat{q}_{c}^{case}
\hat{\sigma}_{i,c}
\right].
\end{equation}

\subsection{Smoothness regularization for residual scale fields}
\label{subsec:smoothness_regularization}

For surface-field prediction, the residual-scale field is expected to be locally coherent on the vehicle surface. Therefore, a local-neighborhood smoothness regularization is introduced to reduce nonphysical point-wise oscillations.

Let \(\mathcal{E}\) denote the set of local neighborhood edges, constructed from a \(k_{\mathrm{nn}}\)-nearest-neighbor graph on the surface points. A subset of surface points may be sampled for efficiency. The smoothness loss is defined as
\begin{equation}
\mathcal{L}_{smooth}
=
\frac{1}{|\mathcal{E}|}
\sum_{(i,j)\in\mathcal{E}}
\left\|
\log
\left(
\hat{\boldsymbol{\sigma}}_{i}
+
\epsilon
\right)
-
\log
\left(
\hat{\boldsymbol{\sigma}}_{j}
+
\epsilon
\right)
\right\|_{2}^{2}.
\end{equation}
This loss encourages neighboring surface points to have similar log-scale estimates. It is applied only to surface-field residual-scale estimation and is not used for scalar drag coefficient prediction.

\subsection{Cross-validation-assisted out-of-fold conformal prediction}
\label{subsec:cv_assisted_calibration}

To improve data utilization and assess the sensitivity of conformal intervals to data partitioning, this work evaluates a CV-assisted out-of-fold (OOF) calibration strategy. The available training-calibration data are divided into \(K\) folds. For each fold \(k\), a surrogate model is trained on the remaining \(K-1\) folds and evaluated on the held-out fold. The resulting held-out predictions provide OOF calibration scores, which are pooled to estimate the conformal quantile. In final evaluation, this pooled calibration information is applied to the corresponding test predictions, as specified in Section~\ref{sec:experimental_setup}.

Let \(\mathcal{I}_k\) denote the index set of the held-out fold \(k\). For each \(j\in\mathcal{I}_k\), the fold model produces an OOF prediction. These OOF predictions are then used to compute calibration scores. The pooled OOF score set is written as
\begin{equation}
\mathcal{S}_{OOF}
=
\bigcup_{k=1}^{K}
\left\{
s_j^{(k)}
:
j\in\mathcal{I}_k
\right\}.
\end{equation}
The corresponding conformal quantile is computed from the pooled OOF scores:
\begin{equation}
\hat{q}_{OOF}
=
Q_{1-\alpha}
\left(
\mathcal{S}_{OOF}
\right).
\end{equation}
This OOF quantile is then used to calibrate predictions on the test set.

For scalar drag coefficient prediction, the OOF predictions contain the lower and upper quantile estimates \(\hat{q}_{0.05}^{(k)}(\mathcal{X}_j)\) and \(\hat{q}_{0.95}^{(k)}(\mathcal{X}_j)\). The asymmetric lower and upper scores are pooled over all held-out folds:
\begin{equation}
\mathcal{S}_{L}^{OOF}
=
\bigcup_{k=1}^{K}
\left\{
\max
\left[
\hat{q}_{0.05}^{(k)}
\left(
\mathcal{X}_j
\right)
-
C_{d,j},
0
\right]
:
j\in\mathcal{I}_k
\right\},
\end{equation}
\begin{equation}
\mathcal{S}_{U}^{OOF}
=
\bigcup_{k=1}^{K}
\left\{
\max
\left[
C_{d,j}
-
\hat{q}_{0.95}^{(k)}
\left(
\mathcal{X}_j
\right),
0
\right]
:
j\in\mathcal{I}_k
\right\}.
\end{equation}
The OOF asymmetric calibration offsets are then computed as
\begin{equation}
\hat{q}_{L}^{OOF}
=
Q_{1-\alpha/2}
\left(
\mathcal{S}_{L}^{OOF}
\right),
\qquad
\hat{q}_{U}^{OOF}
=
Q_{1-\alpha/2}
\left(
\mathcal{S}_{U}^{OOF}
\right).
\end{equation}
These offsets are used in the same way as the held-out calibration offsets in Section~\ref{subsec:cd_conformal}.

For surface-field prediction, the OOF predictions contain the mean field, target field, and residual-scale field. For normalized calibration, the OOF residual-normalized scores are pooled component-wise:
\begin{equation}
\mathcal{S}_{c}^{OOF}
=
\bigcup_{k=1}^{K}
\left\{
\frac{
\left|
y_{j,i,c}
-
\hat{y}_{j,i,c}^{(k)}
\right|
}{
\hat{\sigma}_{j,i,c}^{(k)}
+
\epsilon
}
:
j\in\mathcal{I}_k,\ i=1,\ldots,N
\right\}.
\end{equation}
The channel-wise OOF conformal multiplier is then
\begin{equation}
\hat{q}_{c}^{OOF}
=
Q_{1-\alpha}
\left(
\mathcal{S}_{c}^{OOF}
\right).
\end{equation}
This multiplier is used to construct normalized surface-field prediction intervals on the test set.

This procedure is inspired by cross-validation-based conformal methods, but it is used here as OOF score aggregation rather than as a standard cross-validation-plus construction. Its primary role is to improve data utilization under limited CFD-labeled samples, while also reducing the dependence of the calibration quantile on a single train--calibration split.

\section{Experimental Setup}
\label{sec:experimental_setup}

This section describes the dataset, model settings, experimental protocols, evaluation metrics, and implementation details used to evaluate the proposed conformal reliability framework. The experiments cover two prediction settings defined in Section~\ref{subsec:problem_formulation}: case-level \(C_d\) prediction and surface-field prediction of pressure and wall shear stress.

\subsection{Dataset and Experimental Tasks}
\label{subsec:dataset_tasks}

All main experiments are conducted on the DrivAerML dataset~\citep{ashton2024drivaerml}. Under the official DrivAerML split adopted in this work, 400 cases form the training--calibration pool, 34 cases form the validation set, and 50 cases form the official test set. The validation set is used only for training monitoring and model selection; it is excluded from conformal calibration and final test evaluation. The official test set is held out until final evaluation.

The \(C_d\) task is evaluated at the vehicle-case level. Each vehicle geometry is represented by 8192 surface points sampled from the STL mesh, together with global geometry descriptors. Surface pressure and the three wall shear stress components are evaluated channel-wise on sampled surface points. For surface-field training, 8192 surface points are sampled from each training case to balance geometric resolution and computational cost. During calibration, inference, and final evaluation, the trained model is applied to the available inference points of each case, and coverage and interval-width metrics are computed on these points. For the smoothness regularization term, at most 2048 points per case are used to construct local neighborhoods.

A separate random 80/10/10 split is used only for the smoothness ablation, resulting in a 49-case test subset. This ablation isolates the effect of \(\mathcal{L}_{\mathrm{smooth}}\) and is not intended as an OOF comparison. All prediction intervals target 90\% nominal coverage, corresponding to \(\alpha=0.10\).

\subsection{Model Configurations}
\label{subsec:model_configurations}

GeoTransolver is used as the main backbone, and Transolver is used for backbone-transfer experiments. The prediction heads and conformal prediction procedures are described in Sections~\ref{subsec:cd_conformal}--\ref{subsec:smoothness_regularization}. Table~\ref{tab:model_configurations} summarizes the operator configuration, prediction module, and task-specific training setting for each output type.

Common optimizer settings are reported here to keep the table compact. The \(C_d\) models use AdamW with learning rate \(3\times10^{-4}\), weight decay \(10^{-5}\), and batch size 8. The surface-field models use AdamW with learning rate \(3\times10^{-4}\) and a StepLR scheduler with step size 50 and decay factor 0.5. In the Transolver transfer experiments, only the backbone is replaced; the task heads, loss terms, and conformal evaluation workflow follow the corresponding GeoTransolver setting.

For \(C_d\), the 96-dimensional pooled representation is processed by a
256--128--64 MLP to predict the \(0.05\), \(0.50\), and \(0.95\) quantiles.
For the surface-field model, the loss weights are
\(\lambda_{\mathrm{scale}}=0.05\) and
\(\lambda_{\mathrm{smooth}}=0.01\), with \(k_{\mathrm{nn}}=8\).

\begin{table}[htbp]
\centering
\caption{Main task-dependent GeoTransolver configurations.}
\label{tab:model_configurations}

{\footnotesize
\setlength{\tabcolsep}{10pt}
\renewcommand{\arraystretch}{1.15}

\begin{tabular}{lcccc}
\toprule
Output
& Layers
& Hidden dimension
& Heads
& Slices \\
\midrule

\(C_d\)
& 4
& 192
& 4
& 32 \\

\(p,\tau_x,\tau_y,\tau_z\)
& 20
& 256
& 8
& 128 \\

\bottomrule
\end{tabular}
}
\end{table}

\subsection{Experimental Protocols}
\label{subsec:experimental_protocols}

Table~\ref{tab:experimental_groups} summarizes the experimental groups, protocol contrasts, and primary evaluation focus. GeoTransolver is used in the main experiment groups, whereas the final group replaces it with the Transolver backbone for transferability evaluation.

\begin{table}[!ht]
\centering
\caption{Experimental groups and protocol contrasts.}
\label{tab:experimental_groups}

{\footnotesize
\renewcommand{\arraystretch}{1.2}

\begin{tabular*}{\textwidth}{
@{\extracolsep{\fill}}
l c l
@{}}
\toprule
Experiment group
& Output
& Protocol / contrast \\
\midrule

Scalar calibration
& \(C_d\)
& Split CP vs.\ CV-assisted OOF CP \\

Scalar stability
& \(C_d\)
& 500 Monte Carlo resamples \\

Smoothness ablation
& \(p,\tau_x,\tau_y,\tau_z\)
& \(\lambda_{\mathrm{smooth}}=0.01\) vs.\ \(0\) \\

Field calibration
& \(p,\tau_x,\tau_y,\tau_z\)
& Global / point-adaptive / case-wise CP \\

Field OOF comparison
& \(p,\tau_x,\tau_y,\tau_z\)
& Split CP vs.\ CV-assisted OOF CP \\

Backbone transfer
& \(C_d\), fields
& GeoTransolver \(\rightarrow\) Transolver \\

\bottomrule
\end{tabular*}
}
\end{table}

The scalar experiments evaluate case-level coverage, interval width and
score, and resampling stability. The surface-field experiments assess
residual-scale smoothness, channel- and case-level reliability, and the
effect of OOF score pooling. The backbone-transfer experiment evaluates
generalization across neural-operator backbones.

For \(C_d\), the CV-assisted OOF protocol divides the 400 official training cases into five folds. In each fold, 320 cases are used for training and 80 cases for calibration-score computation. The lower- and upper-tail OOF scores from all folds are pooled to obtain \(\hat{q}^{\mathrm{OOF}}_L\) and \(\hat{q}^{\mathrm{OOF}}_U\). A final model is then trained on all 400 training cases and evaluated on the 50 official test cases. The ordinary split CP baseline uses the same five splits but computes conformal offsets from each calibration fold separately, without OOF score pooling or refitting on all 400 cases. A Monte Carlo stability check is performed using 500 random 200/200 calibration--evaluation resamples.

For surface fields, the smoothness ablation compares \(\lambda_{\mathrm{smooth}}=0.01\) with \(\lambda_{\mathrm{smooth}}=0\) under the random 80/10/10 split and is evaluated using point-adaptive normalized CP. The main surface-field experiment uses four folds over the 400 official training cases. In each fold, 300 cases are used for training and 100 cases for calibration-score computation. OOF scores are pooled channel-wise for \(p\), \(\tau_x\), \(\tau_y\), and \(\tau_z\), and a final model is trained on all 400 training cases. The ordinary split CP baseline uses the same four splits but computes conformal quantiles from each calibration fold separately. The three calibration modes are global absolute, point-adaptive normalized, and case-wise normalized CP.

For backbone transfer, Transolver uses the same scalar readout, field branches, loss terms, and conformal workflow as GeoTransolver. The \(C_d\) transfer experiment includes the official-test evaluation and the same 500-resample Monte Carlo check. For surface fields, a single ordinary calibration--test setting is used to avoid repeating the full GeoTransolver experimental matrix while still checking whether the calibration strategies show the same qualitative behavior with a different backbone.

\subsection{Evaluation Metrics}
\label{subsec:evaluation_metrics}

The evaluation emphasizes calibrated interval reliability rather than deterministic point accuracy alone. Deterministic metrics characterize the underlying surrogate predictor, whereas the main uncertainty metrics are empirical coverage, interval width, interval score, and case-level reliability.

For \(C_d\), the point prediction is \(\hat{q}_{0.50}\). Deterministic accuracy is measured by MAE, MSE, RMSE, MAPE, \(R^2\), Pearson correlation, and Spearman correlation. For a test sample \(j\) with interval \(I_j=[L_j,U_j]\) and target \(y_j\), interval quality is measured by
\begin{equation}
\mathrm{Coverage}
=
\frac{1}{n_{\mathrm{test}}}
\sum_{j=1}^{n_{\mathrm{test}}}
\mathbb{I}\{y_j \in [L_j,U_j]\},
\end{equation}
\begin{equation}
\mathrm{Width}
=
\frac{1}{n_{\mathrm{test}}}
\sum_{j=1}^{n_{\mathrm{test}}}
(U_j-L_j),
\end{equation}
and
\begin{equation}
\mathrm{IS}_{\alpha,j}
=
(U_j-L_j)
+
\frac{2}{\alpha}(L_j-y_j)\mathbb{I}\{y_j<L_j\}
+
\frac{2}{\alpha}(y_j-U_j)\mathbb{I}\{y_j>U_j\}.
\end{equation}
The reported interval score is averaged over all test samples.

For surface-field prediction, coverage and width are computed separately for each channel \(c\in\{p,\tau_x,\tau_y,\tau_z\}\):
\begin{equation}
\mathrm{Coverage}_c
=
\frac{1}{BN}
\sum_{b=1}^{B}
\sum_{i=1}^{N}
\mathbb{I}\{y_{b,i,c}\in [L_{b,i,c},U_{b,i,c}]\},
\end{equation}
\begin{equation}
\mathrm{Width}_c
=
\frac{1}{BN}
\sum_{b=1}^{B}
\sum_{i=1}^{N}
(U_{b,i,c}-L_{b,i,c}).
\end{equation}
Here, \(B\) is the number of test cases, \(N\) is the number of inference points per case, and \(c\) denotes the physical channel.

For surface-field prediction, point-pooled coverage is reported as an aggregate engineering metric over all test vehicles and surface points. Because surface points within the same vehicle are spatially correlated, this metric should not be interpreted as coverage over independent test samples or as a sufficient measure of vehicle-level reliability. We therefore also report case-level coverage statistics to assess the reliability of calibrated intervals at the vehicle level.

For surface-field prediction, we also report the number of cases below 90\% coverage and the fifth percentile of case-level coverage. Monte Carlo resampling is used only for \(C_d\), because repeated point-wise resampling of spatially correlated surface fields would overstate the effective sample size.

\subsection{Implementation}
\label{subsec:implementation}

The experiments are implemented in PyTorch using PhysicsNeMo implementations of GeoTransolver and Transolver. Training and inference are executed on NVIDIA A100 GPUs. Surface-field inference outputs are saved as point-wise \texttt{.npz} files containing predictions, targets, and residual-scale estimates. VTP files are generated only for qualitative visualization; quantitative conclusions are based on the saved metric tables.

The conformal calibration step is post-hoc and does not require additional CFD simulations. For scalar \(C_d\), calibration is performed on cached quantile predictions or cached OOF quantile predictions. For surface-field prediction, conformal multipliers are computed from saved prediction, target, and residual-scale arrays. The additional calibration cost is therefore small compared with neural-operator training and CFD data generation.

\section{Results and Discussion}
\label{sec:results}

This section evaluates the proposed conformal prediction framework in terms of interval reliability, interval efficiency, and case-level stability. The primary criterion is whether empirical coverage remains close to the nominal \(90\%\) level. When different methods achieve comparable coverage, narrower intervals are preferred. Deterministic metrics are reported only to characterize the underlying surrogate predictors.

We first analyze scalar \(C_d\) prediction, followed by residual-scale smoothness, multi-granularity surface-field calibration, backbone transferability, and engineering use.

\subsection{Scalar drag coefficient conformal prediction}
\label{subsec:cd_results}

Table~\ref{tab:cd_deterministic_results} reports deterministic \(C_d\) prediction accuracy. In the ordinary split baseline, five split CP runs are performed on the same folds, with each model trained on 320 cases. In the OOF protocol, fold models generate out-of-fold calibration scores, while final test predictions are produced by a model refitted on all 400 official training cases. This full-data refit explains the lower OOF prediction errors: MAE decreases from 5.44 to 4.97, RMSE decreases from 7.45 to 6.49, and \(R^2\) increases from 0.823 to 0.869.

\begin{table}[htbp]
\centering
\caption{Deterministic point-prediction results for scalar drag coefficient prediction. MAE and RMSE are reported in units of \(10^{-3}\), MSE in \(10^{-5}\), and MAPE in percent.}
\label{tab:cd_deterministic_results}
\footnotesize
\setlength{\tabcolsep}{4pt}
\renewcommand{\arraystretch}{1.15}

\begin{tabular*}{\textwidth}{
@{\extracolsep{\fill}}
lccccccc
@{}}
\toprule
Method
& MAE
& MSE
& RMSE
& MAPE
& \(R^2\)
& Pearson
& Spearman \\
\midrule

Split CP
& 5.44
& 5.70
& 7.45
& 1.91
& 0.823
& 0.911
& 0.901 \\

OOF CP
& 4.97
& 4.21
& 6.49
& 1.76
& 0.869
& 0.941
& 0.940 \\

\bottomrule
\end{tabular*}
\end{table}

Table~\ref{tab:cd_interval_results} reports interval quality on the official test set. The uncalibrated \(q_{0.05}\)--\(q_{0.95}\) interval is included to assess whether conformal calibration is needed. For ordinary split CP, the values are averaged over the five independent split CP runs. For OOF CP, the conformal offsets are computed from the pooled OOF scores and applied to the final full-data model.

The uncalibrated intervals under-cover the nominal \(90\%\) level in both protocols, with 82.40\% coverage for split CP and 84.00\% for OOF CP. After conformal calibration, coverage increases to 88.40\% and 94.00\%, respectively. This confirms that the raw quantile intervals are not automatically calibrated. It also supports the expansion-only calibration design, in which conformal correction repairs under-coverage rather than shrinking the initial interval. The calibrated OOF interval is slightly wider than the calibrated split interval, but it achieves higher coverage and a lower interval score.

\begin{table}[htbp]
\centering
\caption{Uncalibrated and calibrated interval quality for scalar drag coefficient prediction on the official test set. Width and IS are reported in units of \(10^{-3}\).}
\label{tab:cd_interval_results}

{\footnotesize
\setlength{\tabcolsep}{5pt}
\renewcommand{\arraystretch}{1.15}

\begin{tabular*}{\textwidth}{
@{\extracolsep{\fill}}
llccc
@{}}
\toprule
Method
& Interval type
& Coverage (\%)
& Width
& IS \\
\midrule

Split CP
& Uncalibrated PI90
& 82.40
& 18.70
& 32.38 \\

Split CP
& Calibrated CP
& 88.40
& 21.20
& 31.11 \\

OOF CP
& Uncalibrated PI90
& 84.00
& 18.10
& 26.16 \\

OOF CP
& Calibrated CP
& 94.00
& 22.70
& 26.81 \\

\bottomrule
\end{tabular*}
}
\end{table}

The Monte Carlo resampling results in Table~\ref{tab:cd_mc_results} further assess scalar interval stability. In each resampling trial, conformal offsets are recomputed from cached predictions on a randomly selected calibration subset and evaluated on a disjoint evaluation subset; no model retraining is performed. This design isolates the variability caused by calibration--evaluation resampling rather than training randomness. 

Both methods achieve mean coverage close to the nominal level, with 89.71\% for split CP and 89.85\% for OOF CP. Their stability differs, however. Split CP gives a smaller mean width, but its coverage standard deviation is 10.41 percentage points. OOF CP reduces this value to 3.10 percentage points and also slightly reduces width variability. The pooled OOF scores therefore make the calibrated coverage less sensitive to calibration--evaluation resampling. The larger mean width of OOF CP is the cost of this improved stability.

\begin{table}[htbp]
\centering
\caption{Monte Carlo resampling results for scalar drag coefficient conformal prediction. Width statistics are reported in units of \(10^{-3}\), and pp denotes percentage points.}
\label{tab:cd_mc_results}

{\footnotesize
\setlength{\tabcolsep}{4pt}
\renewcommand{\arraystretch}{1.15}

\begin{tabular*}{\textwidth}{
@{\extracolsep{\fill}}
lcccc
@{}}
\toprule
Method
& Mean coverage (\%)
& Std. coverage (pp)
& Mean width
& Std. width \\
\midrule

Split CP
& 89.71
& 10.41
& 25.00
& 2.50 \\

OOF CP
& 89.85
& 3.10
& 31.60
& 1.90 \\

\bottomrule
\end{tabular*}
}
\end{table}

A calibration-size sensitivity analysis using cached OOF predictions is provided in ~\ref{app:cd_calibration_size_sensitivity}. Increasing the calibration size reduces the variability of both empirical coverage and interval width, while maintaining average coverage close to the nominal \(90\%\) level.

Overall, the scalar \(C_d\) results show that the CV-assisted OOF protocol improves data utilization and enables a stronger full-data predictor, while the pooled OOF scores provide substantially greater coverage stability under Monte Carlo resampling.

\subsection{Residual-scale smoothness ablation for surface-field prediction}
\label{subsec:smoothness_ablation_results}

The smoothness ablation tests whether local-neighborhood regularization improves the residual-scale field without degrading conformal coverage. This is important because, in the residual-normalized surface-field CP strategies, the calibrated interval width is scaled by the learned residual scale \(\hat{\sigma}\). Fragmented or isolated variations in \(\hat{\sigma}\) can therefore make the resulting uncertainty maps difficult to interpret, even when the mean prediction is accurate. 

In this ablation, point-adaptive normalized CP is used as the evaluation mode because it directly reflects the spatial behavior of \(\hat{\sigma}\). The regularized model uses \(\lambda_{\mathrm{smooth}}=0.01\), whereas the baseline uses \(\lambda_{\mathrm{smooth}}=0\). All other model and calibration settings are unchanged.

For quantitative evaluation, a channel-wise smoothness score is computed using the same local-neighborhood criterion as the training regularizer:
\begin{equation}
S_c
=
\frac{1}{|\mathcal{E}|}
\sum_{(i,j)\in\mathcal{E}}
\left[
\log\left(\hat{\sigma}_{i,c}+\epsilon\right)
-
\log\left(\hat{\sigma}_{j,c}+\epsilon\right)
\right]^2,
\label{eq:test_smoothness_score}
\end{equation}
where \(c\in\{p,\tau_x,\tau_y,\tau_z\}\). The overall score is the mean of the four channel-wise scores. A lower score indicates smaller local variation and therefore a smoother residual-scale field. The score is evaluated on the 49-case random-split test subset using \(k_{\mathrm{nn}}=8\) nearest neighbors and at most 2048 sampled points per case. As a supplementary check, we also evaluated the deterministic point predictions of the models with and without smoothness regularization. The resulting pressure and wall-shear-stress errors changed only marginally, showing that the smoothness term does not noticeably compromise point prediction accuracy.

Table~\ref{tab:smooth_interval_ablation} reports point-adaptive normalized conformal prediction results. Coverage remains close to the target level for all four channels. The largest rounded coverage change is only 0.15 percentage points, indicating that smoothness regularization does not materially affect empirical coverage.

At comparable coverage, the regularized model produces consistently narrower intervals. Mean width decreases by \(6.12\%\) for pressure, \(5.52\%\) for \(\WSS_x\), \(8.32\%\) for \(\WSS_y\), and \(8.05\%\) for \(\WSS_z\). The improvement is therefore not caused by wider intervals. Instead, smoother residual-scale estimates produce more efficient point-adaptive conformal intervals.

\begin{table}[htbp]
\centering
\caption{Effect of smoothness regularization on point-adaptive normalized conformal prediction. Coverage and width reduction are reported in percent, coverage changes in percentage points (pp), and widths in units of \(10^{-3}\).}
\label{tab:smooth_interval_ablation}

{\footnotesize
\setlength{\tabcolsep}{3pt}
\renewcommand{\arraystretch}{1.15}

\begin{tabular*}{\textwidth}{
@{\extracolsep{\fill}}
lcccccc
@{}}
\toprule

& \multicolumn{3}{c}{Coverage}
& \multicolumn{3}{c}{Width} \\

\cmidrule(lr){2-4}
\cmidrule(lr){5-7}

\shortstack[l]{Physical\\quantity}
& \shortstack{Without\\smoothness}
& \shortstack{With\\smoothness}
& \shortstack{Change\\(pp)}
& \shortstack{Without\\smoothness}
& \shortstack{With\\smoothness}
& \shortstack{Reduction\\(\%)} \\

\midrule

Pressure
& 89.41
& 89.41
& \(-0.01\)
& 58.60
& 55.00
& 6.12 \\

\(\WSS_x\)
& 89.65
& 89.65
& \(-0.01\)
& 0.55
& 0.52
& 5.52 \\

\(\WSS_y\)
& 89.55
& 89.70
& \(+0.15\)
& 0.42
& 0.38
& 8.32 \\

\(\WSS_z\)
& 89.69
& 89.69
& \(+0.01\)
& 0.37
& 0.34
& 8.05 \\

\bottomrule
\end{tabular*}
}
\end{table}

The direct smoothness evaluation in Table~\ref{tab:smooth_score_ablation} confirms this regularization effect. The mean pressure smoothness score decreases from \(0.256\) to \(0.071\), corresponding to a \(72.15\%\) reduction. The reductions for \(\WSS_x\), \(\WSS_y\), and \(\WSS_z\) are \(74.57\%\), \(75.65\%\), and \(73.66\%\), respectively. The overall four-channel score decreases from \(0.465\) to \(0.120\), a reduction of \(74.29\%\). For every output channel, the score ranges of the two models do not overlap, showing that the improvement is consistent across test cases rather than driven by a few selected samples.

\begin{table}[htbp]
\centering
\caption{Channel-wise smoothness scores of the predicted residual-scale fields. Values are reported as the mean and standard deviation over 49 test cases. Lower values indicate smoother fields.}
\label{tab:smooth_score_ablation}

{\footnotesize
\setlength{\tabcolsep}{3pt}
\renewcommand{\arraystretch}{1.15}

\begin{tabular*}{\textwidth}{
@{\extracolsep{\fill}}
lccccc
@{}}
\toprule

& \multicolumn{2}{c}{Mean \(\pm\) std.}
& \multicolumn{2}{c}{Range}
& \\

\cmidrule(lr){2-3}
\cmidrule(lr){4-5}

\shortstack[l]{Physical\\quantity}
& \shortstack{Without\\smoothness}
& \shortstack{With\\smoothness}
& \shortstack{Without\\smoothness}
& \shortstack{With\\smoothness}
& \shortstack{Reduction\\(\%)} \\

\midrule

Pressure
& \(0.256 \pm 0.021\)
& \(0.071 \pm 0.006\)
& \([0.215,\ 0.304]\)
& \([0.058,\ 0.087]\)
& 72.15 \\

\(\WSS_x\)
& \(0.448 \pm 0.021\)
& \(0.114 \pm 0.005\)
& \([0.410,\ 0.501]\)
& \([0.101,\ 0.129]\)
& 74.57 \\

\(\WSS_y\)
& \(0.583 \pm 0.034\)
& \(0.142 \pm 0.009\)
& \([0.511,\ 0.654]\)
& \([0.123,\ 0.159]\)
& 75.65 \\

\(\WSS_z\)
& \(0.574 \pm 0.031\)
& \(0.151 \pm 0.008\)
& \([0.515,\ 0.638]\)
& \([0.136,\ 0.170]\)
& 73.66 \\

\midrule

Overall
& \(0.465 \pm 0.023\)
& \(0.120 \pm 0.006\)
& \([0.414,\ 0.514]\)
& \([0.107,\ 0.136]\)
& 74.29 \\

\bottomrule
\end{tabular*}
}
\end{table}

\begin{figure}[H]
    \centering
    \includegraphics[width=0.88\linewidth]
    {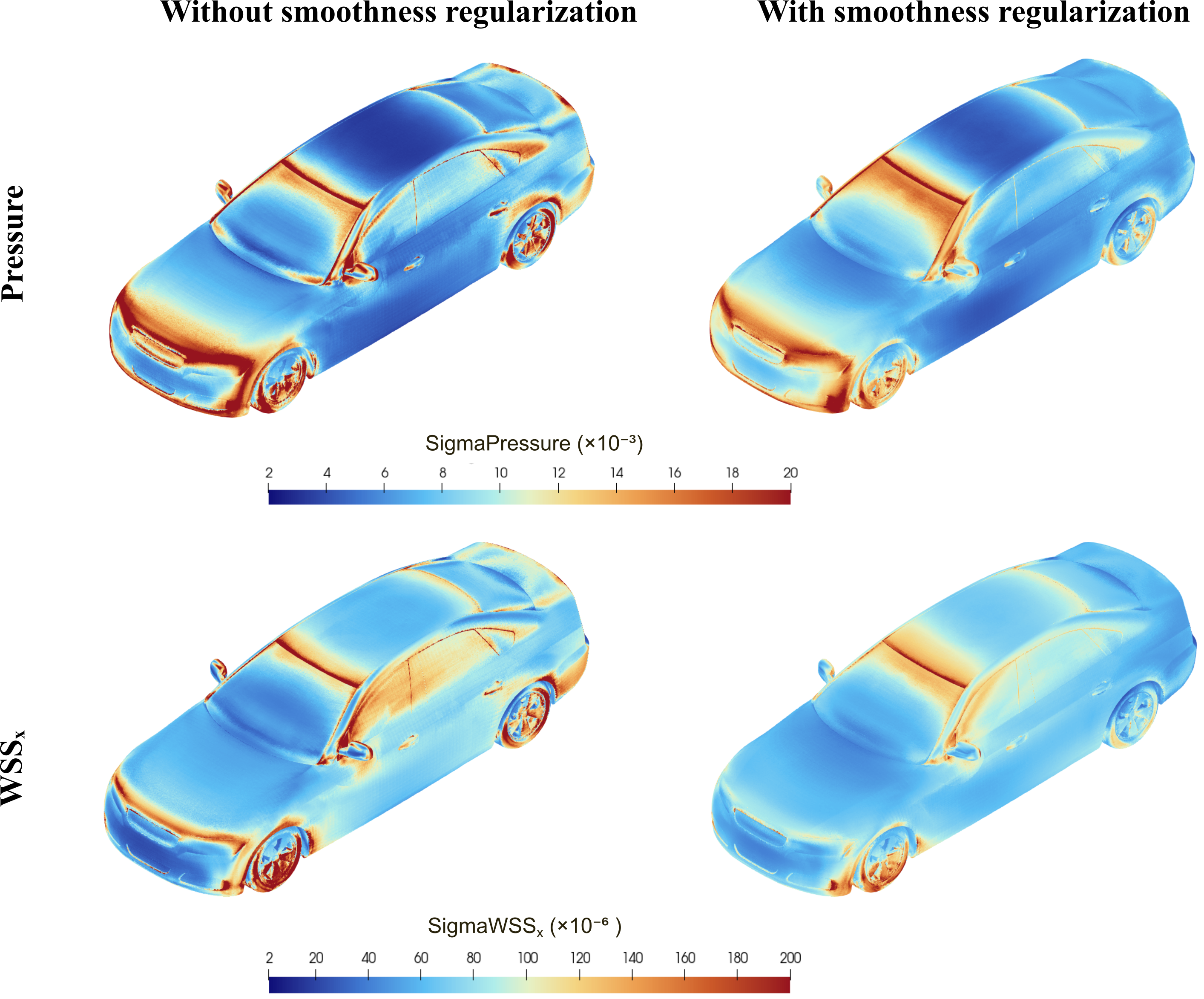}
    \caption{Predicted residual-scale fields for a representative vehicle.
    The left and right columns show the results without and with smoothness
    regularization, respectively, while the top and bottom rows correspond
    to the pressure and \(\mathrm{WSS}_x\) channels.}
    \label{fig:sigma_smooth_visualization}
\end{figure}

Figure~\ref{fig:sigma_smooth_visualization} shows the pressure and
\(\WSS_x\) residual-scale fields for a representative test vehicle.
A normalized residual-scale (\(\sigma\)) head is used because
\(\hat{\sigma}\) provides the local scaling factor in the
residual-normalized conformal score, allowing the interval width to adapt
to spatial variations in prediction error. These fields therefore
visualize the spatial structure of the locally adaptive uncertainty
estimate. A four-channel comparison for one representative test vehicle
is provided in ~\ref{app:smoothness_visualization}.

Without smoothness regularization, the residual-scale fields contain localized high-value streaks and fragmented variations near the windshield boundary, side mirrors, wheels, and sharp body edges. After regularization, isolated variations are substantially reduced, and the residual-scale fields become more spatially coherent. Elevated uncertainty remains visible around geometrically complex regions, indicating that the regularizer does not simply force a uniform field. 

Overall, the smoothness ablation shows that the local-neighborhood regularizer improves residual-scale smoothness and reduces conformal interval width while preserving nearly the same empirical coverage. Based on these results, smoothness regularization is used as the default configuration in the subsequent multi-granularity surface-field experiments.

\subsection{Multi-granularity surface-field conformal prediction}
\label{subsec:surface_multigranularity_results}

\begin{figure}[htbp]
    \centering
    \includegraphics[width=\textwidth]
    {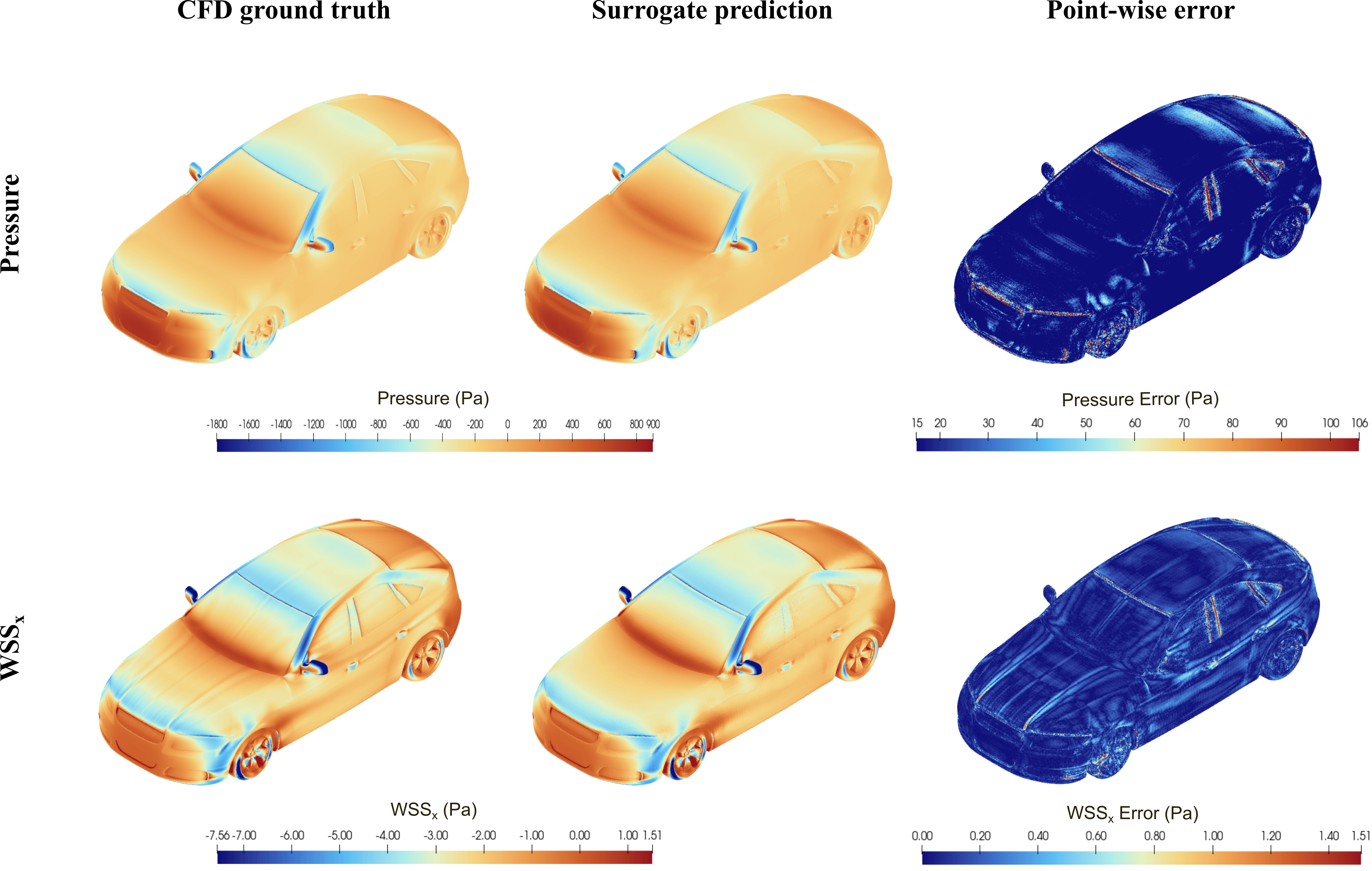}
    \caption{Representative deterministic surface-field predictions on a test vehicle.
    The top row shows the relative pressure field, while the bottom row shows the
    streamwise wall-shear-stress component, WSS\(_x\). The columns present the
    CFD ground truth, the GeoTransolver surrogate prediction, and the corresponding
    point-wise prediction error.}
    \label{fig:deterministic_pressure_wss_prediction}
\end{figure}

This subsection evaluates the three surface-field calibration strategies defined in Section~\ref{subsec:calibration_granularity}: global absolute calibration, point-adaptive normalized calibration, and case-wise normalized calibration. To provide visual context before the quantitative comparison, Fig.~\ref{fig:deterministic_pressure_wss_prediction} shows a representative surface-field prediction example, including the CFD-derived target field, the surrogate prediction, and the point-wise error on the same vehicle surface. This figure is not intended to assess GeoTransolver as a new architecture; rather, it helps connect the subsequent conformal calibration results to the aerodynamic field quantities being calibrated.

The comparison is conducted under the ordinary split CP and CV-assisted OOF protocols. Within each protocol, the three strategies use the same point predictions and differ only in how conformal intervals are constructed.

Table~\ref{tab:surface_deterministic_results} reports deterministic surface-field accuracy. The CV-assisted OOF protocol gives lower errors and higher correlations for all channels, mainly because the final predictor is trained on all 400 official training cases rather than 300 cases in each ordinary split model. The following analysis therefore focuses on interval coverage, width, and case-level reliability.

\begin{table}[htbp]
\centering
\caption{Deterministic surface-field prediction accuracy. Metrics are computed for each test case and channel first, and then averaged over test cases. Relative \(L_1\) error is computed as \(\sum |\hat{y}-y|/(\sum |y|+\epsilon)\).}
\label{tab:surface_deterministic_results}

{\footnotesize
\setlength{\tabcolsep}{5.2pt}
\renewcommand{\arraystretch}{1.12}

\begin{tabular*}{\textwidth}{
@{\extracolsep{\fill}}
llccc
@{}}
\toprule
Channel
& Protocol
& Relative \(L_1\) error (\%)
& \(R^2\)
& Pearson \\
\midrule

Pressure
& Ordinary split
& 6.39
& 0.986
& 0.993 \\

Pressure
& CV-assisted OOF
& 5.38
& 0.990
& 0.995 \\

\(\WSS_x\)
& Ordinary split
& 11.88
& 0.980
& 0.990 \\

\(\WSS_x\)
& CV-assisted OOF
& 10.14
& 0.985
& 0.992 \\

\(\WSS_y\)
& Ordinary split
& 20.54
& 0.971
& 0.985 \\

\(\WSS_y\)
& CV-assisted OOF
& 17.44
& 0.979
& 0.989 \\

\(\WSS_z\)
& Ordinary split
& 19.03
& 0.965
& 0.982 \\

\(\WSS_z\)
& CV-assisted OOF
& 16.28
& 0.973
& 0.986 \\

\bottomrule
\end{tabular*}
}
\end{table}

Table~\ref{tab:surface_overall_interval_quality} compares overall coverage and interval width. Under the ordinary split protocol, global absolute CP and point-adaptive normalized CP both achieve coverage close to the nominal \(90\%\) level. Relative to global absolute CP, point-adaptive normalized CP gives narrower intervals, reducing mean width by \(9.15\%\) for pressure and by \(12.72\%\)--\(14.97\%\) for WSS. Case-wise normalized CP is more conservative, raising coverage to 91.67\%--92.69\%, with a small pressure-width increase but reduced WSS widths.

Under the CV-assisted OOF protocol, the differences become clearer. Global absolute CP becomes conservative, with coverage between 92.24\% and 92.90\%, but its intervals are spatially uniform. Point-adaptive normalized CP gives the narrowest near-nominal intervals, with coverage between 90.15\% and 90.34\% and width reductions of \(22.68\%\) for pressure and \(25.35\%\)--\(27.09\%\) for WSS relative to global absolute CP. Case-wise normalized CP lies between the two, giving higher coverage than point-adaptive normalized CP but wider intervals.

\begin{table}[htbp]
\centering
\caption{Overall coverage and interval width for surface-field conformal prediction. Width change is measured relative to global absolute CP within the same protocol and channel.}
\label{tab:surface_overall_interval_quality}

{\footnotesize
\setlength{\tabcolsep}{2pt}
\renewcommand{\arraystretch}{1.12}

\begin{tabular*}{\textwidth}{
@{\extracolsep{\fill}}
llccc
@{}}
\toprule
\shortstack[l]{Physical\\channel}
& \shortstack[l]{Calibration\\method}
& \shortstack{Overall coverage\\(\%)}
& \shortstack{Mean width\\\((10^{-3})\)}
& \shortstack{Width change\\(\%)} \\
\midrule

\multicolumn{5}{l}{\textbf{Panel A: Ordinary split CP}} \\
\midrule
Pressure & Global absolute CP & 90.25 & 50.40 & -- \\
Pressure & Point-adaptive normalized CP & 90.14 & 45.80 & $-9.15$ \\
Pressure & Case-wise normalized CP & 92.69 & 51.90 & $+2.83$ \\

\(\WSS_x\) & Global absolute CP & 90.36 & 0.50 & -- \\
\(\WSS_x\) & Point-adaptive normalized CP & 90.34 & 0.44 & $-12.72$ \\
\(\WSS_x\) & Case-wise normalized CP & 92.05 & 0.48 & $-4.48$ \\

\(\WSS_y\) & Global absolute CP & 90.04 & 0.37 & -- \\
\(\WSS_y\) & Point-adaptive normalized CP & 90.21 & 0.32 & $-13.76$ \\
\(\WSS_y\) & Case-wise normalized CP & 91.67 & 0.34 & $-6.76$ \\

\(\WSS_z\) & Global absolute CP & 90.15 & 0.34 & -- \\
\(\WSS_z\) & Point-adaptive normalized CP & 90.29 & 0.29 & $-14.97$ \\
\(\WSS_z\) & Case-wise normalized CP & 92.21 & 0.32 & $-5.83$ \\

\midrule
\multicolumn{5}{l}{\textbf{Panel B: CV-assisted OOF protocol}} \\
\midrule
Pressure & Global absolute CP & 92.90 & 50.10 & -- \\
Pressure & Point-adaptive normalized CP & 90.27 & 38.70 & $-22.68$ \\
Pressure & Case-wise normalized CP & 92.79 & 43.70 & $-12.71$ \\

\(\WSS_x\) & Global absolute CP & 92.61 & 0.50 & -- \\
\(\WSS_x\) & Point-adaptive normalized CP & 90.34 & 0.37 & $-25.35$ \\
\(\WSS_x\) & Case-wise normalized CP & 92.15 & 0.41 & $-17.83$ \\

\(\WSS_y\) & Global absolute CP & 92.29 & 0.37 & -- \\
\(\WSS_y\) & Point-adaptive normalized CP & 90.27 & 0.27 & $-25.94$ \\
\(\WSS_y\) & Case-wise normalized CP & 91.76 & 0.29 & $-19.80$ \\

\(\WSS_z\) & Global absolute CP & 92.24 & 0.33 & -- \\
\(\WSS_z\) & Point-adaptive normalized CP & 90.15 & 0.24 & $-27.09$ \\
\(\WSS_z\) & Case-wise normalized CP & 91.98 & 0.27 & $-19.76$ \\
\bottomrule
\end{tabular*}
}
\end{table}

Table~\ref{tab:surface_case_level_stability} further evaluates coverage at the vehicle-case level. 
Under the ordinary split protocol, case-wise normalized CP reduces the number of cases below \(90\%\) coverage for every channel. 
For pressure, the count decreases from 19/50 under both global absolute CP and point-adaptive normalized CP to 4/50 under case-wise normalized CP. 
For \(\WSS_y\), it decreases from 21/50 and 17/50 to 3/50, respectively.

Under the CV-assisted OOF protocol, global absolute CP has the fewest low-coverage cases because it is the most conservative strategy. 
Point-adaptive normalized CP gives the narrowest intervals but less stable case-level coverage, whereas case-wise normalized CP preserves spatial adaptivity while reducing low-coverage cases relative to point-adaptive normalized CP.

\begin{table}[htbp]
\centering
\caption{Case-level coverage stability for surface-field conformal prediction. The fifth percentile (\(P_5\)) is computed from the distribution of case-level coverage over the 50 test vehicles. Standard deviations are reported in percentage points (pp).}
\label{tab:surface_case_level_stability}

{\footnotesize
\setlength{\tabcolsep}{2pt}
\renewcommand{\arraystretch}{1.12}

\begin{tabular*}{\textwidth}{
@{\extracolsep{\fill}}
llcccc
@{}}
\toprule
& & & \multicolumn{3}{c}{Case-level coverage} \\
\cmidrule(lr){4-6}

\shortstack[l]{Physical\\channel}
& \shortstack[l]{Calibration\\method}
& \shortstack{Cases below\\90\%}
& \shortstack{Mean\\(\%)}
& \shortstack{Std.\\(pp)}
& \shortstack{\(P_5\)\\(\%)} \\
\midrule

\multicolumn{6}{l}{\textbf{Panel A: Ordinary split CP}} \\
\midrule
Pressure & Global absolute CP & 19/50 & 90.22 & 1.47 & 88.00 \\
Pressure & Point-adaptive normalized CP & 19/50 & 90.11 & 2.40 & 84.71 \\
Pressure & Case-wise normalized CP & 4/50 & 92.67 & 1.88 & 88.29 \\

\(\WSS_x\) & Global absolute CP & 12/50 & 90.34 & 1.16 & 88.13 \\
\(\WSS_x\) & Point-adaptive normalized CP & 16/50 & 90.32 & 1.53 & 86.97 \\
\(\WSS_x\) & Case-wise normalized CP & 4/50 & 92.03 & 1.33 & 89.12 \\

\(\WSS_y\) & Global absolute CP & 21/50 & 90.06 & 0.90 & 88.57 \\
\(\WSS_y\) & Point-adaptive normalized CP & 17/50 & 90.23 & 1.11 & 88.03 \\
\(\WSS_y\) & Case-wise normalized CP & 3/50 & 91.69 & 0.98 & 89.76 \\

\(\WSS_z\) & Global absolute CP & 19/50 & 90.17 & 1.01 & 88.50 \\
\(\WSS_z\) & Point-adaptive normalized CP & 16/50 & 90.29 & 1.33 & 87.54 \\
\(\WSS_z\) & Case-wise normalized CP & 3/50 & 92.20 & 1.13 & 89.84 \\

\midrule
\multicolumn{6}{l}{\textbf{Panel B: CV-assisted OOF protocol}} \\
\midrule
Pressure & Global absolute CP & 2/50 & 92.87 & 1.34 & 90.69 \\
Pressure & Point-adaptive normalized CP & 15/50 & 90.24 & 2.71 & 84.09 \\
Pressure & Case-wise normalized CP & 6/50 & 92.76 & 2.16 & 87.82 \\

\(\WSS_x\) & Global absolute CP & 2/50 & 92.59 & 1.05 & 90.53 \\
\(\WSS_x\) & Point-adaptive normalized CP & 15/50 & 90.32 & 1.68 & 86.71 \\
\(\WSS_x\) & Case-wise normalized CP & 4/50 & 92.13 & 1.46 & 89.03 \\

\(\WSS_y\) & Global absolute CP & 1/50 & 92.30 & 0.83 & 91.24 \\
\(\WSS_y\) & Point-adaptive normalized CP & 17/50 & 90.28 & 1.23 & 87.83 \\
\(\WSS_y\) & Case-wise normalized CP & 4/50 & 91.77 & 1.09 & 89.62 \\

\(\WSS_z\) & Global absolute CP & 1/50 & 92.25 & 0.95 & 90.74 \\
\(\WSS_z\) & Point-adaptive normalized CP & 19/50 & 90.15 & 1.44 & 87.07 \\
\(\WSS_z\) & Case-wise normalized CP & 4/50 & 91.97 & 1.25 & 89.25 \\
\bottomrule
\end{tabular*}
}
\end{table}

\begin{figure}[H]
    \centering
    \includegraphics[width=\linewidth]{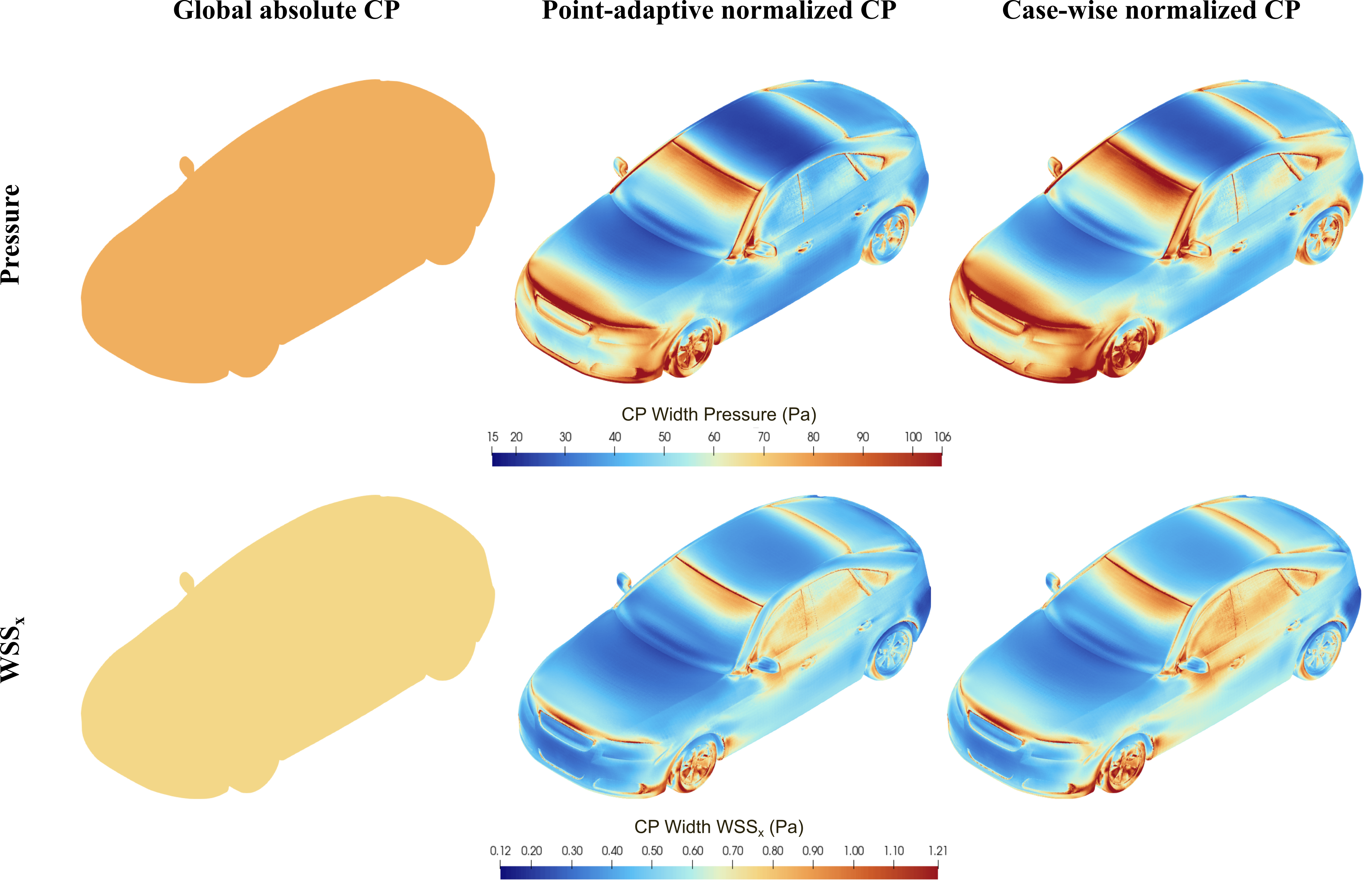}
    \caption{Spatial distribution of conformal interval width under the CV-assisted OOF protocol. Rows show pressure and \(\WSS_x\), and columns compare global absolute, point-adaptive normalized, and case-wise normalized calibration. The same viewing configuration and scalar range are used within each row.}
    \label{fig:surface_interval_width_visualization}
\end{figure}

Figure~\ref{fig:surface_interval_width_visualization} shows the spatial behavior behind these differences. Global absolute CP assigns a uniform width within each channel and cannot adapt to local error variation. In contrast, the two normalized strategies produce spatially varying intervals through the learned residual scale. Point-adaptive normalized CP gives the most compact intervals. Qualitatively, regions with larger interval widths tend to overlap with areas of larger point-wise prediction error in the representative pressure example, suggesting that the learned residual scale captures part of the spatial error heterogeneity. Case-wise normalized CP uses larger calibrated multipliers to improve empirical vehicle-level coverage stability. A full four-channel comparison of the interval-width fields is provided in ~\ref{app:four_channel_interval_width}. 

Overall, the multi-granularity comparison shows a trade-off between interval efficiency and case-level stability. Point-adaptive normalized CP is the most efficient strategy in terms of interval width. Case-wise normalized CP is more conservative, but it improves empirical vehicle-level reliability. Global absolute CP is simple and stable when conservative, but it lacks spatial adaptivity.

\subsection{Backbone transferability from GeoTransolver to Transolver}
\label{subsec:backbone_transfer_results}

This subsection examines whether the proposed conformal prediction pipeline transfers from GeoTransolver to another neural-operator backbone. Transolver is used as the transfer backbone for both scalar drag coefficient and surface-field prediction. The experiment is intended as a transferability check rather than an architecture-optimization study; the task heads, loss terms, and calibration procedures are therefore kept unchanged.

For scalar drag coefficient prediction, Transolver uses the same scalar readout and quantile-regression setting as the GeoTransolver \(C_d\) model. Table~\ref{tab:transolver_cd_transfer} reports deterministic prediction accuracy and calibrated interval quality on the official test set. In the scaled units used for the \(C_d\) results, Transolver achieves an MAE of 5.70, an RMSE of 7.88, and an \(R^2\) value of 0.807. After asymmetric conformal calibration, the interval covers 96.00\% of test cases with a mean width of \(25.70\times10^{-3}\). This conservative result shows that the scalar calibration procedure remains applicable after replacing the backbone.

\begin{table}[htbp]
\centering
\caption{Transolver transfer results for scalar drag coefficient prediction on the official test set. MAE, RMSE, interval width, and IS are reported in units of \(10^{-3}\), MSE in \(10^{-5}\), and MAPE and coverage in percent.}
\label{tab:transolver_cd_transfer}

{\footnotesize
\setlength{\tabcolsep}{3pt}
\renewcommand{\arraystretch}{1.12}

\begin{tabular*}{\textwidth}{
@{\extracolsep{\fill}}
lccccccc
@{}}
\toprule
Model
& MAE
& MSE
& RMSE
& MAPE
& \(R^2\)
& Pearson
& Spearman \\
\midrule

Transolver
& 5.70
& 6.22
& 7.88
& 1.99
& 0.807
& 0.914
& 0.918 \\

\bottomrule
\end{tabular*}

\vspace{0.8em}

\begin{tabular*}{0.88\textwidth}{
@{\extracolsep{\fill}}
lccc
@{}}
\toprule
Model
& Coverage
& Width
& IS \\
\midrule

Transolver
& 96.00
& 25.70
& 32.86 \\

\bottomrule
\end{tabular*}
}
\end{table}

Because the official scalar test set contains only 50 samples, a Monte Carlo resampling experiment is further conducted to evaluate calibration stability. Table~\ref{tab:transolver_cd_mc_transfer} reports the resulting calibrated coverage and tail-error statistics. The mean calibrated coverage is 90.11\%, close to the nominal 90.00\% target. The mean lower- and upper-tail error rates are 4.90\% and 4.99\%, respectively, both close to the nominal 5.00\% target. These results show that scalar conformal calibration remains well balanced under repeated resampling after backbone replacement.

\begin{table}[htbp]
\centering
\caption{Monte Carlo resampling results for Transolver scalar drag coefficient conformal calibration. Each trial uses 200 calibration samples and 200 evaluation samples. Standard deviations are reported in percentage points (pp).}
\label{tab:transolver_cd_mc_transfer}
\footnotesize
\setlength{\tabcolsep}{5.0pt}
\renewcommand{\arraystretch}{1.12}
\begin{tabular*}{0.82\textwidth}{@{\extracolsep{\fill}}lccc}
\toprule
Metric & Target (\%) & Mean (\%) & Std. (pp) \\
\midrule
Calibrated coverage & 90.00 & 90.11 & 2.95 \\
Below calibrated lower bound & 5.00 & 4.90 & 2.13 \\
Above calibrated upper bound & 5.00 & 4.99 & 2.17 \\
\bottomrule
\end{tabular*}
\end{table}

For surface-field prediction, the same three calibration strategies are evaluated: global absolute CP, point-adaptive normalized CP, and case-wise normalized CP. Only the backbone is replaced by Transolver. The residual-scale branch, smoothness loss, and conformal evaluation workflow are kept unchanged. The calibration uses 100 calibration cases, and evaluation is conducted on the 50 official test cases.

Table~\ref{tab:transolver_surface_transfer} reports the surface-field transfer results for pressure and the three wall-shear-stress components. Global absolute CP gives point-pooled coverage close to the 90\% target, but its interval width is spatially uniform within each channel. Point-adaptive normalized CP maintains near-nominal coverage while reducing mean width by \(22.97\%\) for pressure, \(14.57\%\) for \(\WSS_x\), \(18.60\%\) for \(\WSS_y\), and \(12.53\%\) for \(\WSS_z\), relative to global absolute CP. Case-wise normalized CP gives higher coverage, ranging from 91.41\% to 92.57\%, and reduces the number of cases below 90\% coverage to 5/50 for pressure, 4/50 for \(\WSS_x\), 1/50 for \(\WSS_y\), and 2/50 for \(\WSS_z\). Compared with point-adaptive normalized CP, it is more conservative; compared with global absolute CP, it still reduces the mean interval width for all four channels.

\begin{table}[!htbp]
\centering
\caption{Transolver transfer results for surface pressure and wall shear stress conformal prediction. Width change is measured relative to global absolute CP within the same channel. \(P_5\) denotes the fifth percentile of case-level coverage.}
\label{tab:transolver_surface_transfer}

{\scriptsize
\setlength{\tabcolsep}{1pt}
\renewcommand{\arraystretch}{1.12}

\begin{tabular*}{\textwidth}{
@{\extracolsep{\fill}}
llccccc
@{}}
\toprule
\shortstack[l]{Physical\\channel}
& \shortstack[l]{Calibration\\method}
& \shortstack{Coverage\\(\%)}
& \shortstack{Mean width\\\((10^{-3})\)}
& \shortstack{Width change\\(\%)}
& \shortstack{Cases below\\90\%}
& \shortstack{\(P_5\)\\(\%)} \\
\midrule
Pressure
& Global absolute CP
& 89.87
& 461.90
& --
& 29/50
& 88.70 \\
Pressure
& Point-adaptive normalized CP
& 90.02
& 355.70
& $-22.97$
& 17/50
& 87.15 \\
Pressure
& Case-wise normalized CP
& 92.11
& 384.50
& $-16.76$
& 5/50
& 89.75 \\

\(\WSS_x\)
& Global absolute CP
& 90.42
& 2.93
& --
& 10/50
& 89.18 \\
\(\WSS_x\)
& Point-adaptive normalized CP
& 90.82
& 2.50
& $-14.57$
& 10/50
& 87.62 \\
\(\WSS_x\)
& Case-wise normalized CP
& 92.57
& 2.72
& $-7.06$
& 4/50
& 89.68 \\

\(\WSS_y\)
& Global absolute CP
& 89.99
& 1.88
& --
& 25/50
& 89.22 \\
\(\WSS_y\)
& Point-adaptive normalized CP
& 90.35
& 1.53
& $-18.60$
& 16/50
& 88.67 \\
\(\WSS_y\)
& Case-wise normalized CP
& 91.64
& 1.65
& $-12.17$
& 1/50
& 90.07 \\

\(\WSS_z\)
& Global absolute CP
& 89.67
& 1.70
& --
& 33/50
& 88.16 \\
\(\WSS_z\)
& Point-adaptive normalized CP
& 90.12
& 1.49
& $-12.53$
& 20/50
& 88.62 \\
\(\WSS_z\)
& Case-wise normalized CP
& 91.41
& 1.60
& $-6.27$
& 2/50
& 90.03 \\
\bottomrule
\end{tabular*}
}
\end{table}

Overall, the Transolver results show that the calibration pipeline is not tied to GeoTransolver. Scalar intervals remain calibrated under repeated resampling, and residual-scale-based normalized calibration continues to reduce surface-field interval width while preserving near-nominal coverage. These findings indicate that the framework can transfer across neural-operator backbones, provided that the surrogate supplies the outputs required by each calibration mode.

\subsection{Engineering use of calibrated uncertainty}
\label{subsec:engineering_use}

The calibrated intervals are intended for reliability-aware screening of candidate vehicle geometries before high-fidelity CFD results are available. For scalar drag coefficient prediction, the interval
\[
\mathcal{I}_{C_d}=[L_{C_d},U_{C_d}]
\]
defines the calibrated range of the unknown \(C_d\) value. If a design target \(C_d^{target}\) lies inside this interval,
\[
L_{C_d} \leq C_d^{target} \leq U_{C_d},
\]
the surrogate alone cannot confidently determine whether the candidate satisfies the target. Such ambiguous cases, as well as cases with large \(C_d\) interval widths, can be prioritized for follow-up CFD verification.

For surface pressure and wall shear stress, the interval width
\[
w_{i,c}=U_{i,c}-L_{i,c}
\]
acts as a local uncertainty map over the vehicle surface. Large widths indicate regions where the surrogate prediction is less reliable, such as sharp body edges, wheels, underbody structures, or areas with strong local flow variation. A case-level surface uncertainty score can also be obtained from high-percentile interval widths, for example the 95th percentile over surface points.

The proposed framework therefore does not replace high-fidelity CFD. Instead, it provides calibrated reliability indicators for deciding which vehicle geometries and surface regions should be prioritized when CFD resources are limited.

A limitation of the current framework is that conformal coverage relies on the exchangeability between calibration and deployment samples. Therefore, the calibrated intervals should not be interpreted as a standalone solution to out-of-distribution (OOD) generalization. If a new vehicle geometry is far from the training--calibration distribution, the empirical coverage observed on the official test set may no longer hold. In such cases, the intervals should be interpreted cautiously as reliability-screening indicators rather than as distribution-free coverage guarantees, and additional CFD verification remains necessary.

\section{Conclusion}
\label{sec:conclusion}

This work developed a conformal-prediction framework for neural-operator-based automotive aerodynamic surrogate modeling on the DrivAerML dataset. The framework covers both scalar drag coefficient prediction and surface-field prediction of pressure and wall shear stress. GeoTransolver was used as the main backbone, while Transolver was used to examine transferability across neural-operator architectures.

For scalar \(C_d\) prediction, conformalized quantile regression produced calibrated intervals with coverage close to the nominal \(90\%\) level. The CV-assisted OOF protocol improved data use and reduced coverage variability under Monte Carlo resampling. For surface-field prediction, the learned residual-scale branch enabled spatially adaptive intervals. Smoothness regularization improved the coherence of the residual-scale field, while preserving coverage and reducing interval width.

The multi-granularity comparison showed a consistent trade-off between interval efficiency and vehicle-level reliability. Point-adaptive normalized CP produced the narrowest intervals. Case-wise normalized CP was more conservative but reduced the number of low-coverage vehicle cases. Global absolute CP provided a simple channel-wise baseline, but it lacked spatial adaptivity.

The Transolver experiments showed that the calibration pipeline is not tied to GeoTransolver, supporting transfer across neural-operator backbones. Overall, the framework turns deterministic aerodynamic surrogate predictions into calibrated reliability indicators for early-stage design screening. Since conformal coverage relies on exchangeability, future work will combine calibrated conformal intervals with applicability-domain or OOD scores. Such scores could quantify how far a candidate geometry is from the training--calibration distribution and be used together with conformal interval width to form a more complete uncertainty indicator for distribution-shift settings. This direction can further support active CFD sampling by prioritizing geometries that are both uncertain and far from the learned data distribution.

\section*{CRediT authorship contribution statement}
\textbf{Chundong Jia:} Conceptualization, Investigation, Methodology, Data
processing, Writing – original draft. \textbf{Chao Xia:} Conceptualization, Methodology, Writing – review
\& editing, Funding acquisition, Supervision. \textbf{Alexey Vdovin:} Writing – review \&
editing. \textbf{Qing Jia:} Supervision. \textbf{Simone Sebben:} Writing – review
\& editing, Funding acquisition. \textbf{Zhigang Yang:} Supervision.

\section*{Declaration of competing interest}
The authors declare that they have no known competing financial interests or personal relationships that could have appeared to
influence the work reported in this paper.

\section*{Data and code availability}

The DrivAerML dataset used in this study is publicly available from the
original dataset source. The implementation of the conformal calibration,
evaluation, and visualization workflow is available at the following
repository:

\begin{sloppypar}
\noindent\url{https://github.com/DevinJia19/Multi-Granularity-Conformal-Prediction-for-Reliable-Neural-Operator-Automotive-Aerodynamic-Surrogate}
\end{sloppypar}

\section*{Acknowledgements}
This work is funded by the Transport Area of Advance at Chalmers University of Technology through its Seed Project 2026. The computations were enabled by resources provided by the National Academic Infrastructure for Supercomputing in Sweden (NAISS), partially funded by the Swedish Research Council through Grant Agreement No. 2022-06725. 

\clearpage

\appendix

\section{Calibration-size sensitivity for scalar conformal prediction}
\label{app:cd_calibration_size_sensitivity}

This appendix examines how the number of calibration samples affects scalar conformal prediction for \(C_d\). The analysis uses the cached OOF predictions on the 400 official training-calibration cases. No additional model training is performed, and the official test set is not used. For each calibration size, 500 random calibration--evaluation splits are generated, with disjoint calibration and evaluation subsets. This design isolates the effect of calibration-set size from both model-training variability and official test-set evaluation.

Table~\ref{tab:cd_calibration_size_sensitivity} shows that the mean empirical coverage remains close to the nominal \(90\%\) level for all calibration sizes. As the calibration size increases from 40 to 200, the coverage standard deviation decreases from 3.47 to 2.96 percentage points, and the interval-width standard deviation decreases from 4.77 to 1.85 in the scaled units reported in the table. Larger calibration sets therefore provide more stable conformal quantile estimates, consistent with the finite-sample coverage discussion in Section~\ref{subsec:problem_formulation}. The small deviations of mean empirical coverage from exactly \(90\%\) are expected because coverage is estimated on finite evaluation subsets.

\begin{table}[htbp]
\centering
\caption{Calibration-size sensitivity of scalar \(C_d\) conformal prediction using cached OOF predictions. Width statistics are reported in units of \(10^{-3}\), and pp denotes percentage points.}
\label{tab:cd_calibration_size_sensitivity}

{\footnotesize
\setlength{\tabcolsep}{5pt}
\renewcommand{\arraystretch}{1.12}

\begin{tabular*}{0.88\textwidth}{
@{\extracolsep{\fill}}
ccccc
@{}}
\toprule
\shortstack{Calibration\\size}
& \shortstack{Mean\\coverage (\%)}
& \shortstack{Std.\\coverage (pp)}
& \shortstack{Mean\\width}
& \shortstack{Std.\\width} \\
\midrule

40
& 91.21
& 3.47
& 34.66
& 4.77 \\

80
& 90.69
& 3.32
& 33.04
& 3.36 \\

120
& 90.22
& 3.05
& 32.31
& 2.61 \\

160
& 90.31
& 2.97
& 32.01
& 2.23 \\

200
& 89.95
& 2.96
& 31.66
& 1.85 \\

\bottomrule
\end{tabular*}
}
\end{table}

\section{Additional residual-scale visualizations}
\label{app:smoothness_visualization}

This appendix provides a four-channel qualitative comparison of the predicted residual-scale fields for one representative test vehicle. The results obtained without and with smoothness regularization are shown using the same camera position, color map, and scalar range within each channel.

\begin{figure}[htbp]
    \centering
    \includegraphics[
        width=\linewidth,
        height=0.9\textheight,
        keepaspectratio
    ]{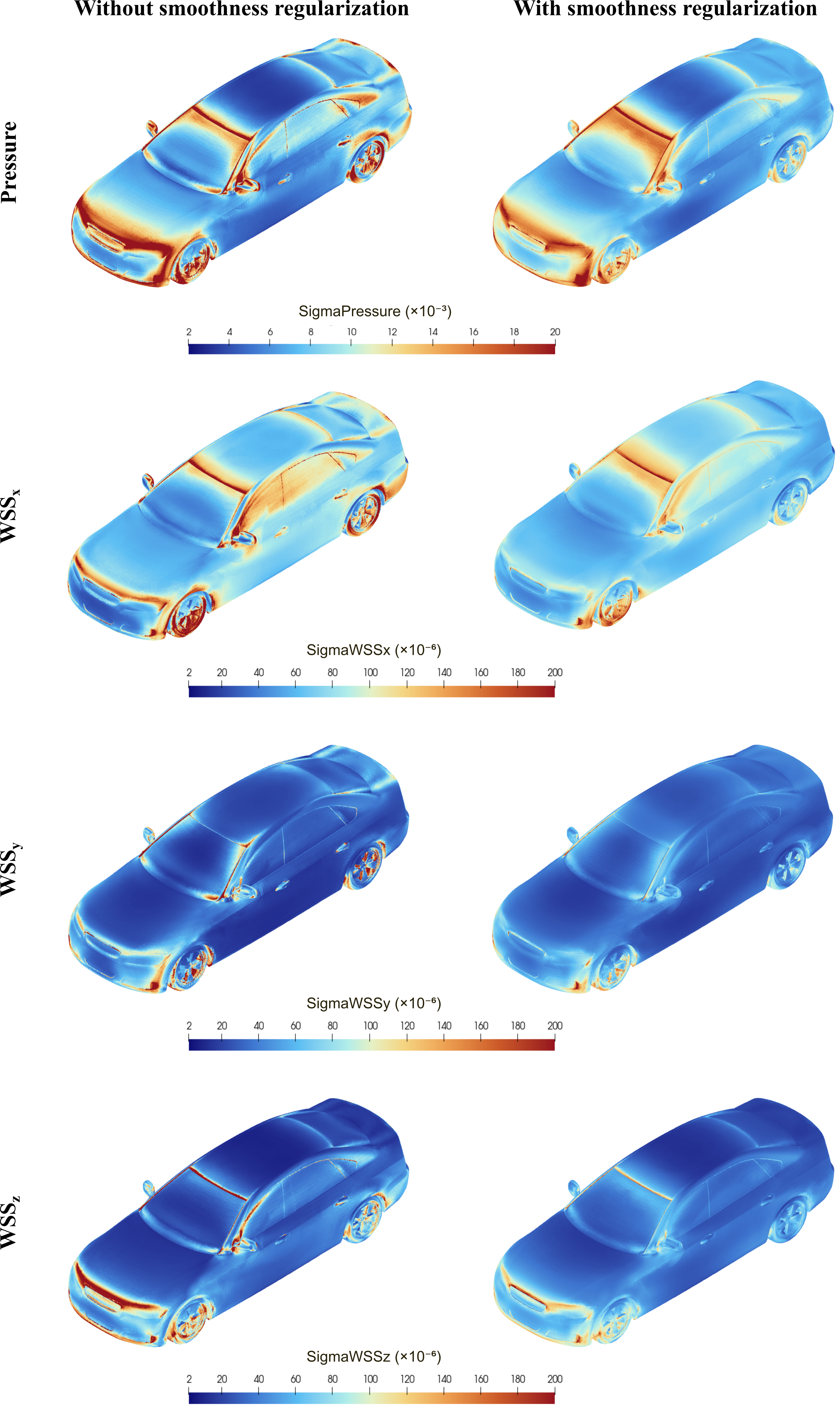}

    \caption{Residual-scale fields for one representative test vehicle.
    Rows correspond to pressure and the three wall-shear-stress components,
    while columns compare the models without and with smoothness regularization.
    The same viewing configuration and scalar range are used within each row.}
    \label{fig:appendix_sigma_four_channels}
\end{figure}

\section{Four-channel visualization of conformal interval width}
\label{app:four_channel_interval_width}

\begin{figure}[H]
    \centering
    \includegraphics[
        width=\linewidth,
        height=0.68\textheight,
        keepaspectratio
    ]{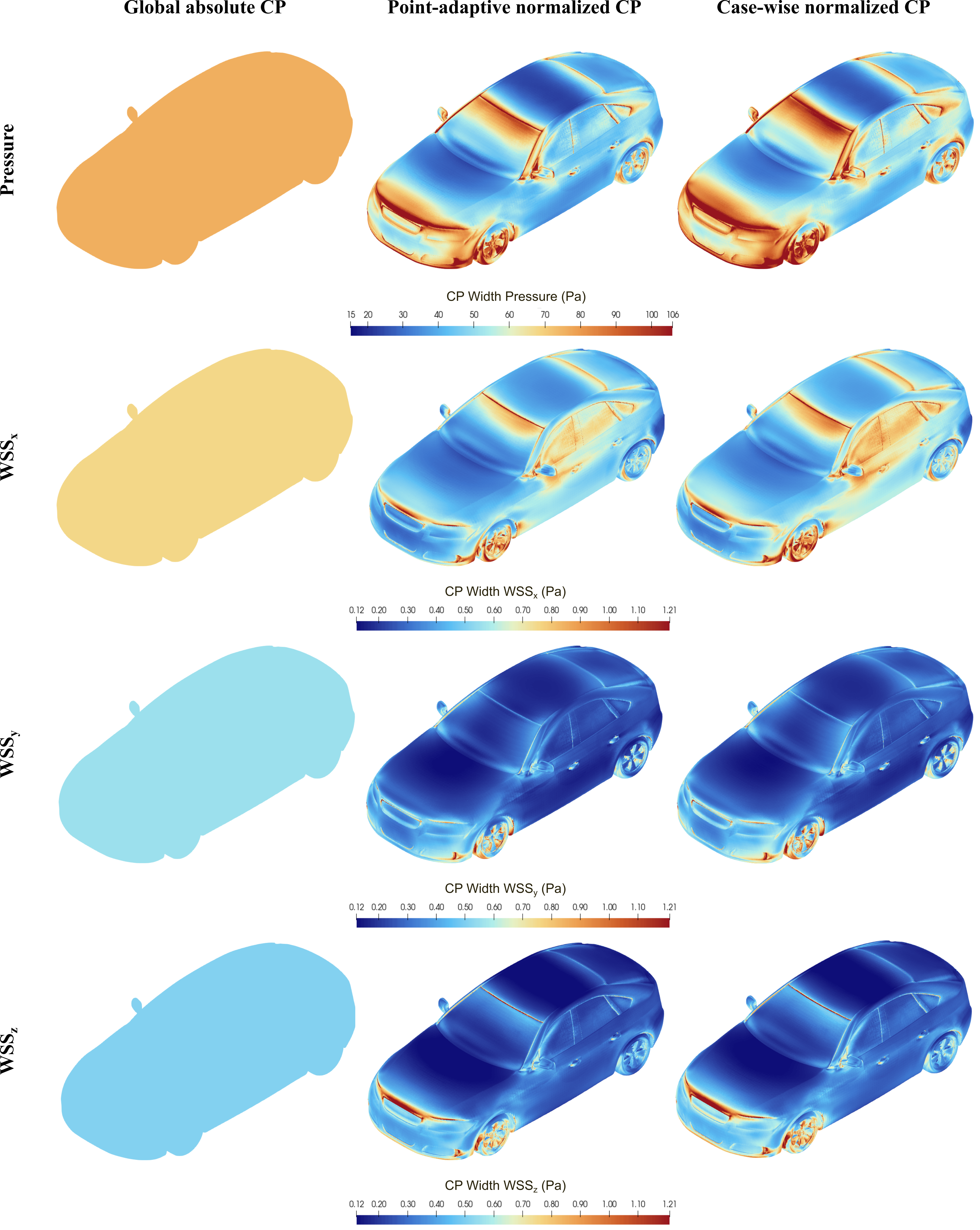} 

    \caption{Four-channel spatial distribution of conformal interval width under the CV-assisted OOF protocol. Rows correspond to pressure, \(\WSS_x\), \(\WSS_y\), and \(\WSS_z\), while columns compare global absolute, point-adaptive normalized, and case-wise normalized calibration. The same viewing configuration and scalar range are used within each row.}
    \label{fig:appendix_four_channel_width}
\end{figure}

Figure~\ref{fig:appendix_four_channel_width} provides the four-channel visualization of conformal interval width under the CV-assisted OOF protocol. The rows correspond to pressure and the three wall-shear-stress components, while the columns compare global absolute, point-adaptive normalized, and case-wise normalized calibration. The same viewing configuration and scalar range are used within each row.

\clearpage

\bibliographystyle{elsarticle-harv}
\bibliography{cas-refs}

\end{document}